\documentclass[a4paper,fleqn]{cas-sc}

\usepackage[authoryear]{natbib}

\usepackage[justification=centering]{caption}
\usepackage{subcaption}
\usepackage{placeins}

\usepackage{amsmath,amsfonts}
\usepackage{bm}
\usepackage{xcolor}
\usepackage{booktabs}

\DeclareMathOperator*{\argmin}{arg\,min}

\graphicspath{{figs/}}

\newcommand{\newtextcolor}{black}

\newcommand{\parder}[2]{\frac{\partial #1}{\partial #2}}

\newcommand{\dt}{{\Delta t}}
\newcommand{\dx}{{\Delta x}}

\usepackage{framed} 
\usepackage{multicol} 

\usepackage{nomencl} 
\makenomenclature
\renewcommand\nomgroup[1]{%
  \ifthenelse{\equal{#1}{A}}{%
    \item[\textbf{Acronyms}]}{
  \ifthenelse{\equal{#1}{R}}{%
    \item[\textbf{Roman Symbols}]}{
  \ifthenelse{\equal{#1}{G}}{%
    \item[\textbf{Greek Symbols}]}{
 {
  \ifthenelse{\equal{#1}{S}}{%
    \item[\textbf{Subscripts/superscripts}]}{
  \ifthenelse{\equal{#1}{X}}{%
    \item[\textbf{Other Symbols}]}{
  {}}}}}}}}
  
\renewcommand*\nompreamble{\begin{multicols}{2}}
\renewcommand*\nompostamble{\end{multicols}}

\begin{document}

\nomenclature[r]{$t$}{time}
\nomenclature[r]{$\Delta t$}{time discretization parameter}
\nomenclature[r]{$\Delta x$}{space discretization parameter}
\nomenclature[g]{$\tau$}{relaxation time}
\nomenclature[g]{$\Omega$}{collision operator}
\nomenclature[r]{$g$}{gravitational acceleration}
\nomenclature[g]{$\kappa$}{surface tension adjustment parameter}
\nomenclature[g]{$\delta$}{time shift delay}
\nomenclature[r]{$f$}{LBE distribution functions}
\nomenclature[g]{$\psi$}{pseudopotential}
\nomenclature[g]{$\rho$}{density of the fluid}
\nomenclature[g]{$\gamma$}{surface tension}
\nomenclature[g]{$\sigma$}{thermodynamic consistency parameter}
\nomenclature[r]{$T$}{temperature of the fluid}
\nomenclature[g]{$\eta$}{viscosity of the fluid}
\nomenclature[g]{$\lambda$}{thermal conductivity}
\nomenclature[r]{$c_{\scriptsize\text{v}}$}{constant volume specific heat}
\nomenclature[r]{$\mathbf{u}$}{velocity of the fluid}
\nomenclature[r]{$a,b$}{Peng-Robinson Equation's parameters}
\nomenclature[s]{$\ell$}{liquid}
\nomenclature[s]{$v$}{vapor}
\nomenclature[s]{$w$}{at wall}
\nomenclature[s]{$wh$}{at heater}
\nomenclature[s]{$eq$}{equilibrium}
\nomenclature[s]{$sat$}{saturation}
\nomenclature[s]{$c$}{critical}

\let\WriteBookmarks\relax
\def\floatpagepagefraction{1}
\def\textpagefraction{.001}
\shorttitle{Pseudopotential Lattice Boltzmann Method for boiling heat transfer: a mesh refinement procedure}
\shortauthors{Jaramillo A et~al.}

\title [mode = title]{Pseudopotential Lattice Boltzmann Method for boiling heat transfer: a mesh refinement procedure} 

\author[1]{Alfredo Jaramillo}
\ead{ajaramillopalma@gmail.com}
\cormark[1]

\author[1]{Vinícius Pessoa Mapelli}
\ead{vinicius.mapelli@usp.br}

\address[1]{Department of Mechanical Engineering, Heat Transfer Research Group, São Carlos School of Engineering (EESC), University of São Paulo (USP), São Carlos 13566-590, Brazil}

\author[1]{Luben Cabezas-Gómez}
\ead{lubencg@sc.usp.br}

\cortext[cor1]{Corresponding author}

\begin{abstract}
Boiling is a complex phenomenon where different non-linear physical interactions take place and for which the quantitative modeling of the mechanism involved is not fully developed yet. In the last years, many works have been published focusing on the numerical analysis of this problem. However, a lack of numerical works assessing quantitatively the sensitivity of these numerical simulations to grid parameters can be identified, especially for the Lattice Boltzmann method (LBM). The main goal of this work is to propose a mesh refinement methodology for simulating phase-change heat transfer problems employing the pseudopotential LBM. This methodology was based on relating the physical parameters to their lattice counterparts for an arbitrary mesh under the viscous regime (where $\Delta t \propto \Delta x ^2$). A suitable modification of the EOS parameters, the adjusting of thermodynamic consistency and surface tension for a certain $\Delta x$ were the main steps of the proposed methodology. A first ensemble of simple simulations including the droplet vaporization and the Stefan problems was performed to validate the proposed method and to assess the influence of some physical mechanisms. Global norms in space and time were used to evaluate the variations of both the density and temperature fields for pool boiling simulations when the lattice discretization is refined. It was observed that the proposed methodology provides convergent results for all the problems considered, and the convergence orders depend on the complexity of the simulated phenomena.

\end{abstract}


\maketitle

\section{Introduction}

Boiling heat transfer is a phenomenon widely exploited in common daily tasks as long as in industrial applications, such as nuclear reactors, heat pipes, industrial boilers, refrigeration and power systems, and others (Carey, 2008). \textcolor{\newtextcolor}{In general, pool boiling represents a boiling process at the body surface immersed in a large volume of liquid (Carey, 2008), and it is one of the main boiling mechanisms addressed in the literature. This boiling process takes place in quenching processes, in electronic devices immersed in cooling liquids, and in flooded evaporators employed in industrial refrigeration installations and nuclear power stations using pressurized water reactors, to mention a few applications.} The reason for its widespread adoption is its high efficiency in transferring heat, especially in the nucleate boiling regime, where vapor bubbles cycle through well-known phases of formation, growth, and departure. \textcolor{\newtextcolor}{In this paper, a mesh refinement study for obtaining mesh independent solutions is proposed and applied to three studied problems related to liquid-gas phase-change, namely, the one-dimensional Stefan problem, the evaporation of a two-dimensional droplet in a superheated vapor and the pool boiling problem in two dimensions. The applied methodology is a step forward in the numerical simulation of boiling processes that occur in real applications with the pseudopotential LBM.} 

The mathematical modeling and numerical analysis for this phenomenon represent a challenging problem with inherent non-linear behavior, and physical phenomena evolving along with different scales \citep{Son1999, son2002numerical}. Moreover, solid-liquid-vapor interparticle interaction may take place, which has been proved to have an important influence on quantities of interest like the heat transfer coefficient \citep{10.1115/1.2910737}. For that
reason, several researchers have dedicated themselves to better understand these phenomena, either experimentally, theoretically, or
numerically \citep{Lienhard1973,Haramura1983,Kandlikar2001,Guan2011,Gong_2015_LatticeBoltzmannsimulations,Kim2016,Liang2018,Feng2018,Farinas2019,guzella2020simulation}.

Recently, a numerical method that has drawn attention is the Lattice Boltzmann Method (LBM) 
due to its potential to simulate a variety of complex phenomena such as flow in porous medium \citep{Guo_2002_LatticeBoltzmannmodel, Liu_2016_MultiphaselatticeBoltzmann}, 
turbulence \citep{Yu_2005_DNSLESdecaying, Chen_2004_ExpandedanalogyBoltzmann}, 
particle suspensions \citep{McCullough_2016_LatticeBoltzmannmethods,Safaei_2016_MathematicalModelingNanofluids}
and others \citep{Gong_2012_Numericalinvestigationdroplet,Gong2013,Gong_2015_Numericalsimulationpool,Gong_2015_LatticeBoltzmannsimulations,guzella2020simulation}. {\color{\newtextcolor}A comprehensive review of the method, focusing on multiphase flows, thermal flows, and thermal multiphase flows with phase change has been recently provided by \citet{Li2016}}. One of the LBM advantages for simulating interfacial dynamics lies in the fact that it does not involve front-tracking, as is the case with more traditional methods such as VOF and level-set. LBM can be seen as the discretization of the Boltzmann transport equation so that its relation to kinetic theory is also considered to be an advantage, since it allows to include small and mesoscale particle physical interactions into the mathematical modeling, mimicking the macroscopic flow behavior.


Several LBM models have been developed for multiphase flow simulations \citep{Gunstensen_1991_LatticeBoltzmannmodel,Swift_1995_LatticeBoltzmannSimulation,Luo_1998_Unifiedtheorylattice,ShanChen1993}. Among them, pseudopotential-based models have been the most successful methodologies for simulating phase-change heat transfer phenomena. The original pseudopotential model was proposed by \citet{ShanChen1993}, where 
an interparticle force is responsible for maintaining liquid-vapor phases in equilibrium. 
Since then, many research groups have contributed to its development. \citet{Yuan_2006_Equationsstatelattice} proposed a
pseudopotential method in which a non-ideal equation of states (EOS) could be employed, 
such as the Van Der Waals equation and the Peng-Robinson equation. \citet{Sbragaglia_2007_GeneralizedlatticeBoltzmann} was one of the first to propose a model that allowed to tune the surface tension, a modeling feature that has a key relevance in this work. \citet{Kupershtokh_2009_equationsstatelattice} proposed a 
modified interparticle force by combining the original Shan and Chen method with the one proposed by \citet{Zhang2003}, which resulted in an enhancement of thermodynamic consistency. In the works of \citet{Li2013} and \citet{Li2013b}, the authors 
proposed a consistent approach to adjust surface tension and vapor-liquid coexistence curve. Later, 
\citet{LycettBrown_2015} performed a third-order analysis and argued that latter adjustments,
along with an interface thickness treatment, are required to achieve a consistent grid refinement when the pseudopotential 
method is employed. More recently, \citet{Zhai_2017_PseudopotentiallatticeBoltzmann, Kharmiani_2019_AlternativeHighDensity} and \citet{Czelusniak2020} also proposed consistent models to tune these three aspects of the pseudopotential method.

\begin{table*}[!t]
  \begin{framed}
    \printnomenclature
  \end{framed}
\end{table*}

The pseudopotential method has also been employed in simulations involving phase-change and heat transfer. 
\citet{Gong_2012_Numericalinvestigationdroplet} proposed a similar interparticle force to \citet{Kupershtokh_2009_equationsstatelattice} looking to improve thermodynamical consistency. In a further 
study, \citet{Gong2013} employed a second distribution function to simulate a temperature passive-scalar equation along with the 
aforementioned interparticle force to study nucleation on a microheater under constant wall temperature and constant heat flux conditions. 
Bubble growth and other aspects such as the effects of contact angle, superheat, and gravity were investigated. A reasonable agreement between the simulations results and well-known engineering correlations was observed. Later, \citet{Gong_2015_LatticeBoltzmannsimulations} used that same methodology to investigate pool boiling on mixed wettability surfaces. In their study, it was shown that surfaces with hydrophobic and hydrophilic spots can achieve better heat transfer efficiency by avoiding bubble coalescence. This improvement is also determined by geometry, such as the size and pitch of these spots. In the same year, \citet{Li_2015_LatticeBoltzmannmodeling} employed a model they proposed before \citep{Li2013} along with a finite difference method for simulating the temperature field evolution (called \emph{the hybrid approach}, firstly performed by \citet{Zhang2003}). By increasing the wall superheating, the authors were able to simulate the regimes of nucleation, transition, and film boiling. Later, \cite{Gong_2017_Directnumericalsimulations} 
included some solid nodes on their computational grid to study the heater thermal response under different boiling regimes. 
An analogous configuration has been employed by \citet{Dong2018} to investigate film boiling behavior.
{  \color{\newtextcolor}The influence of an electric field and gravity to enhance the boiling process has been studied by \citet{Feng2019a,Feng2019b}, where the influence of both these force fields on quantities like the Critical Heat Flux was assessed. The same research group, using a Runge-Kutta method to discretize the energy equation, obtained pool boiling curves in different gravity intensities and wall superheats proposing a gravity scaling model to predict heat flux in terms of the superheat and gravity intensity \citet{Feng2021b}}.
Mixed wettability 
surfaces have also been investigated by \citet{Ma_2019_3Dsimulationspool}, with a three-dimensional extension of the model 
proposed by \citet{Gong_2012_Numericalinvestigationdroplet}, {\color{\newtextcolor}a similar study on mixed wettability surfaces was performed by \citet{Feng2021} but using a LBM to solve the energy equation. In the same line, \citet{Li2020a,Li2021} have numerically studied novel heating surfaces designed to enhance boiling heat transfer, with conclusions that have received experimental support from \citet{Sun2022}}. More recently, \citet{Zhao2021} employed a hybrid approach based on \citet{Li2013} to study several aspects of pool boiling on confined and free outlet geometries. {\color{\newtextcolor}Also, \citet{Fei2020} have implemented a three-dimensional cascaded lattice Boltzmann method (CLBM) simulating a complete 3D boiling process, obtaining a good correlation of the critical heat flux predicted by the simulation with established theoretical results. Moreover, it is found that the bubble footprint area distribution changes from an exponential distribution to a power-law distribution when increasing the wall superheat, which is in agreement with experimental research.}

Despite its advantages when applied to phase-change simulations, LBM pseudopotential-based methods could still be considered to be in their early research stage. A great number of works present qualitative results and/or values in the so-called lattice units, which are relative to mesh spacing and sound speed. Even though the conversion to physical units is immediate, their relationship may lead to confusion when a non-ideal EOS is employed. Also, studies showing fundamental aspects of the pseudopotential method are still scarce. For instance, in the work of \citet{Li_2018_temperatureequationphase}, the authors clarify a misconception on the energy equation employed by \citet{Gong_2018_modifiedphasechange}, which has been extensively employed in previous studies. Also, \citet{LycettBrown_2015} and \citet{Zheng2019} show some spurious terms on the macroscopic equations when the forcing scheme proposed by \citet{Kupershtokh_2009_equationsstatelattice} is used, which can also influence the coexistence curve. 

Grid refinement when using the pseudopotential method is not a trivial task, and studies of the grid influence on numerical pool boiling results are scarce. The present work focuses on fundamental numerical aspects of the pseudopotential method when a hybrid approach is used. Specifically, the main goal is to provide a mesh refinement procedure for simulating liquid/vapor phase-change when the pseudopotential-based LBM is employed. This kind of study is essential to perform LBM simulations of multiphase problems for fluids of engineering interest. To attain this goal, the physical parameters were related to their lattice counterparts for an arbitrary mesh under the viscous regime (where $\Delta t \propto \Delta x ^2$) in order to conserve the hydrodynamic and thermodynamic behaviors. The non-trivial part of this task is the suitable modification of the EOS and the adjusting of the surface tension for a certain $\Delta x$. Once these modifications were understood, some simple numerical experiments (the droplet vaporization problem and the Stefan problem) were analyzed numerically to validate the proposed methodology. Lastly, typical pool boiling simulations are analyzed and discussed qualitatively and quantitatively.


The structure of this article is as follows: Section \ref{sec:mathematical-modeling} describes the pseudopotential LBM multiphase model, the contact angle models, and the energy conservation equation employed. In Section \ref{sec:problem-definition}, a mesh refinement procedure to maintain a consistent physical behavior of the fluid between meshes is presented. The grid refinement procedure is validated for the droplet cooling and the Stefan problems in Section \ref{sec:numerical-results}. Pool boiling numerical findings are also shown and discussed in that section. Finally, some concluding remarks are drawn in Section \ref{sec:conclusions}.



\section{Mathematical modelling}
\label{sec:mathematical-modeling}
The LBM is based on a discretized form of the Boltzmann transport equation known as the lattice Boltzmann equation (LBE) \citep{Chen_1998_LatticeBoltzmannmethod,Shan_1998_Discretizationvelocityspace}. The LBE is employed to compute the fluid behavior calculating the evolution of a series of discrete distribution functions $\{f_i\}$ given by:
\begin{equation}
    f_i(\mathbf{x}+ \,\mathbf{c}_i  ,t^{n+1})=f_i(\mathbf{x},t^n)+\Omega_i(\mathbf{f},\mathbf{f}^{\text{eq}})(\mathbf{x},t^n) + \mathbf{R}(\mathbf{x},t^n)~,
    \label{eq:lbm}
\end{equation}


\noindent where ${f}_{\alpha}$ are the particle distribution functions related to
velocity $\mathbf{c}_i$ and 
$f^{\text{eq}}_i$ are the local equilibrium distribution
functions, $\mathbf{f}$, $\mathbf{f}^{\text{eq}}$ correspond to vectors whose components
are $[\mathbf{f}]_{i}$ = $f_{i}$, $[\mathbf{f}^{\text{eq}}]_{i} = f^{\text{eq}}_{i}$,  and $t$ and
$\mathbf{x}$ are time and space coordinates, respectively.
The term $\mathbf{R}$ is a general representation of a source term in the moment space, defined by the forcing scheme employed. Through this term, 
effects of a general external force field $\mathbf{F}$ are taken into account.
The term expressed by $\Omega_i(\mathbf{f},\mathbf{f}^{\text{eq}})$ is the so-called collision operator, 
and it can generally be defined as a quantity dependent on distribution function $\mathbf{f}$ 
and its equilibrium values $\mathbf{f}^{\text{eq}}$.
For the two-dimensional D2Q9 lattice scheme, a finite set of velocities vector $\mathbf{c}_i$ is formulated as \citep{qian1992lattice}:


\begin{equation}
    \mathbf{c}_i=\left\{\begin{array}{ll}
    (0,0)\,c,     i=0~,  \\
      (1,0)\,c, (0,1)\,c, (-1,0)\,c, (0,-1)\,c,   & i=1,\ldots,4~, \\ 
      (1,1)\,c, (-1,1)\,c, (-1,-1)\,c, (1,-1)\,c~,   & i=5,\ldots,8~,
    \end{array}\right.
\end{equation}
where $c = \dx {}/ \dt$.

One of the approaches most commonly employed for the collision operator 
$\Omega_i(\mathbf{f},\mathbf{f}^{\text{eq}})$ was proposed by 
\citet*{Bhatnagar_1954_ModelCollisionProcesses}, and can be written as:
\begin{equation}
    \Omega_i(\mathbf{f},\mathbf{f}^{\text{eq}}) = 
    -\frac{1}{\tau} (f_{i}-f_{i}^{\text{eq}})~,
\end{equation}
where the parameter $\tau$ is the relaxation time. The BGK collision operator is also known as the single-relaxation time operator. A more stable version of this collision operator allows for multiple-relaxation times (MRT) \citep{Higuera_1989_Boltzmannapproachlattice, d1992generalized}, reading:
\begin{equation}
    \Omega_i(\mathbf{f},\mathbf{f}^{\text{eq}})=-\left[\mathbf{M}^{-1}\mathbf{\Lambda}\mathbf{M}\right]_{ij}(f_j-f_j^{\text{eq}})~,
    \label{eq:collisionMRT}
\end{equation}
 $\mathbf{\Lambda}$ being the relaxation matrix
 \begin{equation*}
     \mathbf{\Lambda}=\text{diag}\left(\tau_\rho^{-1},\tau_e^{-1},\tau_\zeta^{-1},\tau_j^{-1},\tau_q^{-1},\tau_j^{-1},\tau_q^{-1},\tau_\eta^{-1},\tau_\eta^{-1}\right)~,
 \end{equation*}
 where $\tau_\rho$ and $\tau_j$ are the relaxation parameters related to density and momentum conserved moments, 
 $\tau_\eta$ defines dynamic viscosity through $\eta = c_s^2 (\tau_\eta - 0.5)\dt$, $\tau_e$ is associated with the bulk viscosity, $\tau_\zeta$ is related to the energy square quantity, and $\tau_q$ is related to the energy flux. The matrix $\mathbf{M}$ is the matrix that converts $f$ into the momentum space, defined as \citep{Lallemand_2000_TheorylatticeBoltzmann}:
\begin{equation*}
    \mathbf{M}=\left(\begin{array}{rrrrrrrrr}
    1 & 1 & 1 & 1 & 1 & 1 & 1 & 1 & 1 \\
    -4 & -1 & -1 & -1 & -1 & 2 & 2 & 2 & 2\\
    4 & -2 & -2 & -2 & -2 & 1 & 1 & 1 & 1 \\
    0 & 1 & 0 & -1 & 0 & 1 & -1 & -1 & 1 \\
    0 & -2 & 0 & 2 & 0 & 1 & -1 & -1 & 1 \\
    0 & 0 & 1 & 0 & -1 & 1 & 1 & -1 & -1 \\
    0 & 0 & -2 & 0 & 2 & 1 & 1 & -1 & -1 \\
    0 & 1 & -1 & 1 & -1 & 0 & 0 & 0 & 0 \\
    0 & 0 & 0 & 0 & 0 & 1 & -1 & 1 & -1 
    \end{array}\right)\,.
\end{equation*}

The equilibrium distribution function values $\mathbf{f}^{\text{eq}}$ can 
be computed through its relation to the macroscopic density ($\rho$) 
and velocity ($\mathbf{u} = u_{\alpha}$) fields as follows:
\begin{equation} 
\label{eq:EquilibriumDistribution}
f_i^{eq} = w_i \bigg( \rho + \frac{c_{i \alpha}}{c_s^2} \rho u_{\alpha}
+ \frac{(c_{i \alpha}c_{i \beta} - c_s^2 \delta_{\alpha \beta})}{2 c_s^4} 
\rho u_{\alpha}u_{\beta} \bigg)~,
\end{equation}	
where the weights $w_i$ are related to each discrete velocity $\mathbf{c}_i$, and $c_s$ is the lattice sound speed. For the D2Q9 velocity set, one has $w_0 = 4{}/9$, $w_1 = w_2 = w_3 = w_4 = 1{}/9$ and $w_5 = w_6 = w_7 = w_8 = 1{}/36$. In that way, the macroscopic density and velocity may be recovered through following relations:
\begin{equation} 
\label{eq:DensityAndVelocityRelationToF}
\rho = \sum_i f_i, \quad \rho \mathbf{u} = \sum_i f_i \mathbf{c}_i + \frac{\mathbf{F}}{2}~.
\end{equation}	

It is worth mentioning that each $f_i$ represents the number of particles placed between $\mathbf{x}$ and $\mathbf{x}+d\mathbf{x}$ at time $t$, and moving with velocity $\mathbf{c}_i$. In the LBM, the spacial ($\Delta x$) and temporal $(\Delta t)$ discrete quantities are commonly fixed to the unity (e.g., 1 m and 1 s when SI units are employed). In Eq. \eqref{eq:lbm} that assumption was considered, implying that $t^{n+1} = t^{n} + 1$ and $ d\mathbf{x} = \mathbf{c}_i\Delta x = \mathbf{c}_i$. 

Therefore, once one has obtained numerical results with a given number of timesteps and lattices distribution, the refinement of these results, to assess mesh sensibility, requires recomputing the non-dimensional parameters that represent the particular physical setting. This issue is further developed in Section \ref{sec:problem-definition}.

\subsection{The pseudopotential method}
The last term from Eq. \eqref{eq:lbm}, $\mathbf{R}$, is the source term, which is described by the forcing scheme employed.
This source term is used for adding effects of a general external force field, $\mathbf{F}$. 
In this work, the enhanced \citet{Li2013} forcing scheme, based on original scheme from \citet{Guo_2002_LatticeBoltzmannmodel}, 
along with \citet{Li2013b} surface tension correction term are employed.
\nomenclature[r]{$G$}{interaction strength}
\nomenclature[r]{$c_s$}{sound's speed}

In this work, the external force field reads $\mathbf{F}=\mathbf{F}^m+\mathbf{F}^{\text{ads}}+\mathbf{F}^b$, 
where  $\mathbf{F}^m$ is the intermolecular force, $\mathbf{F}^\text{ads}$ the adherence force, and $\mathbf{F}^b$ the body force. The intermolecular force field models the bulk two-phase physics through a pseudopotential function, which depends on $\rho$. For single-component fluids, the interaction force may be written \citep{Chen_1998_LatticeBoltzmannmethod}:
\begin{equation}
    \mathbf{F}_i^m(\mathbf{x},t)=-G\,\psi(\mathbf{x},t)\sum_i \text{w}_i \psi(\mathbf{x}+ \mathbf{c}_i,t)\,\mathbf{c}_{i}~,
    \label{eq:shanchen}
\end{equation}
where $\psi$ is the pseudopotential field and $G$ is the interaction strength. 
For the intermolecular force, the weights $\text{w}_i$ are equal to $\text{w}_i = 1{}/3$ for 
$|\mathbf{c}_i|^2 = 1$, and $\text{w}_i = 1{}/12$ when $|\mathbf{c}_i|^2 = 2$ \citep{Sbragaglia_2007_GeneralizedlatticeBoltzmann}.
A prescribed non-ideal EOS can be imposed by using the following relation for the pseudopotential:
\begin{equation}
\psi(\mathbf{x},t)=\sqrt{\frac{2(p_{ \text{\scriptsize EOS}}(\mathbf{x},t)-\rho(\mathbf{x},t)/3)}{G}}~,\label{eq:psi}
\end{equation}
where $p_{ \text{\scriptsize EOS}}$ is the thermodynamic pressure. When using this definition, 
the parameter $G$ no longer represents intermolecular strength, and it is used only to 
assure a positive value inside the square root. The value $G = -1$ is assumed throughout this work.

In this work, the Peng-Robinson EOS is used, which can be written:
\begin{equation}
    p_{ \text{\scriptsize EOS}}=\frac{\rho RT}{1-b\rho}-\frac{a\rho^2\left[1+\left(0.37464+1.54226\,\omega-0.2699\,\omega^2\right)\left(1-\sqrt{T/T_c}\right)\right]^2}{1+2b\rho-b^2\rho^2}~,\label{eq:peng-robinson}
\end{equation}
where $T=T(\mathbf{x},t)$ is the temperature field, $T_c$ is the critical temperature, $\omega$ is the acentric factor (fixed to $0.344$), $R$ is the universal gas constant (fixed to 1), $a=0.45724\,R^2T_c^2/p_c$, $b=0.0778\,R T_c/p_c$ (fixed to $9.52381\times 10^{-2}$ in lattice units), and $p_c$ is the pressure at the critical point. 

Gravitational force effects are incorporated by taking:
\begin{equation}
    \mathbf{F}^{b} = \mathbf{g}\,(\rho - \overline{\rho})~,
    \label{eq:gravitational_force}
\end{equation}
where $\mathbf{g}$ is the gravitational acceleration, and $\overline{\rho}$ is the spatial mean density computed throughout entire 
computational domain at every timestep.

The adherence force $\mathbf{F}^\text{ads}$ may be interpreted as a boundary treatment of the intermolecular force $\mathbf{F}^{m}$. 
The contact angle models considered are better discussed on Section \ref{sec:mathematical-modeling-contact-angle}. The source term $\mathbf{R}$ proposed by \citet{Li2013b} can be defined as:
\begin{equation}
    R_{ij}= \left[\mathbf{M}^{-1}\left(\mathbf{I}-\frac{\mathbf{\Lambda}}{2}\right)\right]_{ij} S_j 
    + \left[\mathbf{M}^{-1}\right]_{ij} C_j~,
    \label{eq:li_source_terms}
\end{equation}
where the forcing term $\mathbf{S}$ is computed by:
\begin{equation}
\mathbf{S}=\begin{pmatrix}
0\\
6(u_x F_x+u_y F_y)+\frac{12\sigma\, (\mathbf{F}^{m}\cdot \mathbf{F}^{m})^2}{\psi^2(\tau_e-0.5)}\\
-6(u_x F_x+ u_y F_y)-\frac{12\sigma\,(\mathbf{F}^{m}\cdot \mathbf{F}^{m})^2}{\psi^2(\tau_\zeta-0.5)}\\
F_x\\
-F_x\\
F_y\\
-F_y\\
2(u_x F_x-u_y F_y)\\
u_x F_y+u_y F_x
\end{pmatrix}~.
    \label{eq:S}
\end{equation}

In Eq. \eqref{eq:S}, the parameter $\sigma$ is used to adjust the thermodynamic consistency, which is specially needed when employing the definition given on Eq. \eqref{eq:psi}
\citep{LycettBrown_2015}. If $\sigma = 0$, first term of Eq. \eqref{eq:li_source_terms} simplifies to 
traditional \citet{Guo_2002_Discretelatticeeffects} forcing scheme.

As for the additional term $\mathbf{C}$, its definition is as follows:
\begin{equation}
    \mathbf{C}=\left[\begin{array}{c}
    0\\
    1.5\tau_e^{-1}\left(Q_{xx}+Q_{yy}\right)\\
    -1.5\tau_\zeta^{-1}\left(Q_{xx}+Q_{yy}\right)\\
    0\\
    0\\
    0\\
    0\\
    -\tau_v^{-1}\left(Q_{xx}-Q_{yy}\right)\\
    -\tau_v^{-1}Q_{xy}
    \end{array}\right]~,\label{eq:surface-tension-correction-term-C}
\end{equation}
with
\begin{equation}
\qquad \mathbf{Q}=\kappa\, \frac{G}{2} \psi(\mathbf{x},t) \sum_i w\left(|\mathbf{c}_i|^2\right) \left[\psi(\mathbf{x}+\mathbf{c}_i,t)-\psi(\mathbf{x},t)\right] \mathbf{c}_i\mathbf{c}_i ~.\label{eq:surface-tension-correction-term-Q}
\end{equation}

The term in moment space $\mathbf{C}$ is a correction term used to tune the desired surface tension through the adjustment of the parameter $\kappa$ present in $\mathbf{Q}$ \citep{Li2013}. 
 
 {\color{\newtextcolor} When applying Chapman-Enskog analysis on described
 method, Navier-Stokes equations in the following form can be derived:
\begin{align} 
\partial_{t} \rho+\nabla \cdot(\rho \mathbf{v}) &=0~, \label{eq:mass-conservation}\\
\partial_{t}(\rho \mathbf{v})+\nabla \cdot(\rho \mathbf{v} \mathbf{v}) &= -\nabla \cdot \mathbf{P} +\nabla \cdot \mathbf{\Pi}+\mathbf{F}^{b} + \mathbf{F}^{ads}~,\label{eq:momentum-conservation}
\end{align}
 
 \noindent where $\mathbf{\Pi} = \rho \eta \left[\nabla \mathbf{u}+(\nabla \mathbf{u})^{T}\right]+\rho(\xi-\eta)(\nabla \cdot \mathbf{u}) \mathbf{I} $, in which $\xi = c_s^2 (\tau_{e} - 0.5)\Delta t$. Pressure tensor is defined by relation $ \nabla \cdot \mathbf{P} = \nabla \cdot \left( \rho c_{s}^{2} \right) - \mathbf{F}^{m}$. In fact, when applying enhanced pseudopotential model proposed 
 by \citet{Li_2013_LatticeBoltzmannmodeling} and \citet{Li_2013_Achievingtunablesurface}, one can rewrite pressure tensor as:
 
 \begin{equation}
\mathbf{P} = \left(\rho c_{s}^{2}+\frac{G c^{2}}{2} \psi^{2}+2\sigma{G^{2} c^{4}}|\nabla \psi|^{2}+\frac{G c^{4}}{12} (1+2 \kappa) \psi \nabla^{2} \psi\right) \mathbf{I} + \frac{G c^{4}}{6} (1 - \kappa) \nabla \psi \nabla \psi~.
\end{equation}
 
The definition of the pressure tensor is relevant when considering static cases, since momentum equation reduces to $\nabla \cdot \mathbf{P} = 0$. For the planar interface, for example, a constant pressure across the domain is expected, and since it is a one dimensional problem, an analytical solution can be found by solving the resulting ordinary differential equation. Also, surface tension can be found by integrating the difference of the pressure tensor components $P_{xx} - P_{yy}$ across the interfacial region. In fact, the interface planar problem allows one to draw important conclusions and comparisons between methods. In this work, we also employ conclusions from the planar interface problem to find suitable values for tuning parameters $\kappa$ and $\sigma$ as the computational domain is refined.
 }

\subsection{Contact angle modeling and surface tension at the solid walls}
\label{sec:mathematical-modeling-contact-angle}

The fluid-solid interaction is given by the Zou-He scheme \citep{Zou1997} and two contact angle models:

\begin{description}
\item[The $\psi$-based model:] In this case, the adherence forces are given by \citep{Benzi2006,Huang2009}
\end{description}
\begin{equation}
    \mathbf{F}^\text{ads}(\mathbf{x},t)=-G\,\psi(\rho(\mathbf{x},t),T(\mathbf{x},t))\,\psi(\rho_w(\mathbf{x}),T^*_w(\mathbf{x},t))\sum_i w_i \,s(\mathbf{x}+\mathbf{c}_{i})\,\mathbf{c}_{i}~,
    \label{eq:psi-based-formula}
\end{equation}

\noindent where $s$ is an indicator function that is equal to the unity only when $\mathbf{x}+\mathbf{c}_{\alpha}$ corresponds to a solid lattice, and $\rho_w(\mathbf{x})$ is an auxiliary function used to model the local solid-fluid interaction.

\begin{description}
\item[The modified-$\psi$-based model:] Under this scheme one imposes \citep{Li2014}:
\end{description}
\begin{equation}
    \mathbf{F}^\text{ads}(\mathbf{x},t)=-G_w(\mathbf{x})\,\psi^2(\rho(\mathbf{x},t),T(\mathbf{x},t))\sum_i w_i\,s(\mathbf{x}+\mathbf{c}_{i})\,\mathbf{c}_{i}~.
    \label{eq:mod-psi-based-formula}
\end{equation}

When using the $\psi$-based scheme, the fluid-solid interaction is modeled through the imposition of a pair of functions $(\rho_w,T^*_w)$ that may depend on both space and time coordinates.

Having selected one of these models for the fluid-solid interaction links, at the wall lattices the Eq. \eqref{eq:shanchen} is replaced by
\begin{equation}
\mathbf{F}_i^m(\mathbf{x},t)=-G\,\psi(\mathbf{x},t)\left(\sum_{j \in I^{\text{fluid}}} w_j \psi(\mathbf{x}+ \mathbf{c}_j ,t)\,c_{ji}\right)+\mathbf{F}_i^\text{ads}(\mathbf{x},t)~,
    \label{eq:shanchen-walls}
\end{equation}
where $I^{\text{fluid}}$ denotes the set of indices for which $\mathbf{x}+ \mathbf{c}_j$ is a fluid lattice.

\subsubsection{Surface tension at the boundary walls}
The correction term defined by Eqs. \eqref{eq:surface-tension-correction-term-C}-\eqref{eq:surface-tension-correction-term-Q} corresponds to discrete approximations of derivatives of the scalar field $\psi$. Thus, its evaluation at wall lattices requires some attention and it is not clear how this calculation has been performed in the literature. In this work, we have calculated this correction term at the boundary walls following the contact angle model computational procedure. More precisely, when the $\psi$-based model is imposed, the term $\mathbf{Q}$ is computed as:
\begin{align}
    \mathbf{Q}&=\kappa\, \frac{G}{2} \psi(\mathbf{x},t) \sum_{i \in I^{\text{fluid}}} w\left(|\mathbf{c}_i|^2\right) \left[\psi(\mathbf{x}+\mathbf{c}_i,t)-\psi(\mathbf{x},t)\right] \mathbf{c}_i\mathbf{c}_i\nonumber\\&+\kappa\, \frac{G}{2} \psi(\mathbf{x},t) \sum_{i \in I^{\text{solid}}} w\left(|\mathbf{c}_i|^2\right) \left[\psi(\rho_w(\mathbf{x},t),T^*_w(\mathbf{x},t))-\psi(\mathbf{x},t)\right] \mathbf{c}_i\mathbf{c}_i ~,\label{eq:Q-psi-based}
\end{align}
analogously, when the modified-$\psi$-based model is used, $Q$ reads
\begin{align}
    \mathbf{Q}&=\kappa\, \frac{G}{2} \psi(\mathbf{x},t) \sum_{i \in I^{\text{fluid}}} w\left(|\mathbf{c}_i|^2\right) \left[\psi(\mathbf{x}+\mathbf{c}_i,t)-\psi(\mathbf{x},t)\right] \mathbf{c}_i\mathbf{c}_i~,\label{eq:Q-mod-psi-based}
\end{align}
which is equivalent to substitute $\psi(\mathbf{x}+\mathbf{c}_i,t)$ by $\psi(\mathbf{x},t)$ in Eq. \eqref{eq:surface-tension-correction-term-Q} when $\mathbf{x}+\mathbf{c}_i$ corresponds to a solid lattice.

\subsection{The energy conservation equation}
The energy conservation equation is solved for computing the scalar temperature field. In the present work, this equation is solved by the Finite Difference (FD) method, following the hybrid approach \citep{Zhang2003,Li_2015_LatticeBoltzmannmodeling}. The evolution of the temperature field, derived from the energy conservation equation, is governed by \citep{Li2020}:
\begin{equation}
		\partial_t T+\mathbf{u}\cdot \nabla T=\frac{1}{\rho c_\text{\scriptsize V}}\nabla\cdot(\lambda \nabla T)-\frac{1}{\rho c_\text{\scriptsize V}}T\left(\parder{p_{\text{\scriptsize EOS}}}{T}\right)_\rho\nabla\cdot \mathbf{u}\label{eq:energy-equation}~,
\end{equation}
with $c_\text{\scriptsize V}$ being the constant volume specific heat, and $\lambda=\lambda(\mathbf{x},t)$ is the thermal conductivity, given by
$$\lambda(\mathbf{x},t) = \frac{\lambda_\ell(\rho(\mathbf{x},t)-\rho_\text{\scriptsize v})+\lambda_\text{\scriptsize v}(\rho_\ell-\rho(\mathbf{x},t))}{\rho_\ell-\rho_\text{\scriptsize v}}~,$$
where $\rho_\ell$ ($\lambda_\ell$) and $\rho_\text{\scriptsize v}$ ($\lambda_\text{\scriptsize v}$) are the density (thermal conductivity) of the liquid and vapor phase respectively. As usual, the viscous dissipation term is assumed to be negligible. For all the simulations performed in this work, we have taken the values of $c_\text{\scriptsize V}$ of both phases as being equal and constant.

Equation \eqref{eq:energy-equation} is discretized through a FD method along the lattices discrete values and weights as described in \cite{Li2018,guzella2020simulation} regardless of the type of boundary condition for $T$ (Dirichlet or heat flux). No special treatment is needed at the boundaries since the \emph{wet-node} approach is used for the LBM \citep{Zou1997}. As for the time derivatives, they are computed according to a second-order Runge-Kutta method adapted from \cite{Leveque2007}. 

\section{Proposed mesh refinement procedure}
\label{sec:problem-definition}

In this section, we present the proposed mesh refinement procedure for simulating liquid/vapor phase-change problems with heat transfer. This procedure will be applied to the solution of the three typical problems presented in Section \ref{sec:numerical-results}. However, the proposed approach can be applied to simulate other phase-change problems with LBM.

 

To refine the grid, it is necessary to choose the set of non-dimensional parameters that define the physical problem. Through this work we use the hat symbol ``\textasciicircum" ~to refer to physical quantities in SI units, otherwise, the quantities are assumed to be expressed in lattice units.  The definition of these non-dimensional parameters is based on working at the viscous regime, considering $\Delta \hat{t}\propto \Delta \hat{x}^2$, where $\Delta \hat{t}$ and $\Delta \hat{x}$ are the physical time step and the physical distance between lattices, respectively. {\color{\newtextcolor} Considering, for sake of simplicity, domains with equal sizes $\hat{L}$ along each direction, in general, the physical mesh size parameter corresponds to the domain size $\hat{L}$ divided by the desired number of lattices ($N$) set to discretize the domain along each direction, $\Delta \hat{x}=\hat{L}/N$}.

\subsection{Physical dependence between meshes}
\label{sec:physics-meshes}

The proposed procedure is addressed to establish the necessary conditions for performing a mesh refinement study to simulate liquid/vapor phase-change problems with heat transfer. The main problem is related to pool boiling, so the gravitational effects are not negligible. For this reason, the chosen physical characteristic length, velocity, and time are those proposed by \citet{Son1999}, respectively:


\begin{equation}
    \hat{\ell}_0=\sqrt{\frac{\hat{\gamma}}{\hat{g}\left(\hat{\rho}_\ell-\hat{\rho}_v\right)}}\,,\qquad \hat{u}_0=\sqrt{\hat{g}\,\hat{\ell}_0}\,,\qquad \hat{t}_0=\hat{\ell}_0/\hat{u}_0 ~,\label{eq:charact-param}
\end{equation}
where $\hat{\ell}_0$ is the capillary length, which represents the ratio between the surface tension and the buoyancy force effects, $\hat{\gamma}$ is the surface tension coefficient, $\hat{g}$ is the gravitational acceleration, and $\hat{\rho}_\ell$ ($\hat{\rho}_v$) is the liquid (vapor) phase density.{ \color{\newtextcolor}In lattice units (l.u.) the characteristic length ($\ell_0$), velocity  ($u_0$), and time ($t_0$) are set as}
\begin{equation}
    \ell_0=\hat{\ell}_0/\Delta \hat{x}\,,\qquad u_0=\sqrt{{g}\,\ell_0}\,,\qquad t_0=\ell_0/u_0 ~,
\end{equation}
where the gravitational acceleration in lattice units is given by $g=\frac{\gamma}{\ell_0^2\left({\rho}_\ell-{\rho}_v\right)}$.

The fluid kinematic viscosity ($\hat{\eta}$) is assumed to be a function of each phase viscosity, liquid $\hat{\eta}_\ell$ and vapor $\hat{\eta}_v$, according to:
\begin{equation}
    \hat{\eta}(\mathbf{x},t)=\frac{\hat{\rho}(\mathbf{x},t)-\hat{\rho}_v}{\hat{\rho}_\ell-\hat{\rho}_v}\,\hat{\eta}_\ell + \frac{\hat{\rho}_\ell -\hat{\rho}(\mathbf{x},t)}{\hat{\rho}_\ell-\hat{\rho}_v}\,\hat{\eta}_v~,
    \label{eq:eta-rho}
\end{equation}
where $\hat{\eta}_\alpha=\hat{u}_0\hat{\ell}_0/\text{Re}_\alpha$, $\alpha\in\{\ell,v\}$. The symbol $\text{Re}_\alpha$ represents the respective Reynolds number of the $\alpha$ phase. Knowing that the physical time step $\Delta \hat{t}$ is related to $\Delta \hat{x}$ by $\Delta \hat{t} = \Delta \hat{x}^2 \eta/\hat{\eta}$ \citep{Kruger2017}, the following relations may be computed:
\begin{equation}
    \Delta \hat{t} = \Delta \hat{x}^2 \frac{\eta}{\hat{\eta}}= \Delta \hat{x}^2 \frac{u_0\,\ell_0}{\hat{u}_0\hat{\ell}_0}~.
\end{equation}

Therefore, to remain in the viscous regime when changing $\Delta \hat{x}$, the quantity $u_0{\ell}_0$ must not depend on $\Delta \hat{x}$. On the other hand, it is possible to compute the relation:
$$u_0{\ell}_0=\sqrt{\frac{\hat{\ell}_0}{\hat{\rho}_\ell-\hat{\rho}_v}\frac{\gamma}{\Delta \hat{x}}}~,$$
and thus one concludes that it is necessary for $\gamma$ to scale as $\mathcal{O}\left(\Delta\hat{x}\right)$. As observed by \cite{Czelusniak2020}, the only way to obtain that behavior is by adjusting the parameter $\kappa$ in Eq. \eqref{eq:surface-tension-correction-term-Q} for each $\Delta\hat{x}$, as the surface tension (computed from the planar interface problem) is proportional to $1-\kappa$. 


Before discussing how this is achieved in this work, let us summarize the physical quantities we want to keep constant among the different meshes. These quantities are chosen in order to ensure the required consistency from similarity laws, mainly considering the hydrodynamic and thermodynamic similarities.  
\begin{itemize}
    \item Hydrodynamic consistency: The Reynolds number $\text{Re}_\alpha$ of each phase, as well as, the surface tension are kept constant. Particularly, the last condition means that the Laplace equation
     \begin{equation}
    \Delta P_{\Delta \hat{x}}= \frac{\gamma}{R_{\Delta\hat{x}}}~,\label{eq:laplace-equation}
    \end{equation}
    must remain valid for any $\Delta\hat{x}$, with $\Delta P_{\Delta \hat{x}}$ being the difference in the pressure field (given by an EOS) inside/outside the bubble, and $R_{\Delta \hat{x}}$ the bubble radius. As we want $\gamma=\mathcal{O}\left(\Delta\hat{x}\right)$ and $R_{\Delta \hat{x}}$ must scale as $\mathcal{O}\left(\Delta\hat{x}^{-1}\right)$ this implies that one must have $\Delta P_{\Delta \hat{x}}=\mathcal{O}\left(\Delta\hat{x}^{2}\right)$.
    \item Thermodynamic consistency: The density ratio is kept constant for a particular imposed saturation temperature ($T_s$) and all meshes. The coexistence curve is adjusted through the parameter $\sigma$ to maintain the same coexistence densities for the $T_s$ imposed. This procedure is not exact, in the sense that the coexistence curve $(T_s,\rho)$ is not exactly the same when changing $\Delta\hat{x}$ \citep{Li2013b}. Due to this reason, the surface tension adjustment through the parameter $\kappa$ needs to be performed for each mesh as a final step of the computation of the parameters.
\end{itemize}

The conversion factor for some arbitrary field $\theta=\theta(\mathbf{x},t)$ is the constant $C_\theta$ such that the physical field $\hat{\theta}(\mathbf{x},t)$ (e.g., expressed in SI units) can be obtained from the lattice units term $\theta(\mathbf{x},t)$ as $\hat{\theta}(\mathbf{x},t)=C_\theta\,\theta(\mathbf{x},t)$ for any pair $(\mathbf{x},t)$. The conversion factors $C_\rho$, $C_p$ and $C_T$ correspond to the critical values $\rho_c$, $p_c$ and $T_c$ (critical density, pressure and temperature, respectively), and they are determined by the EOS in use, which in this work corresponds to the Peng-Robinson EOS, see Eq. \eqref{eq:peng-robinson}. It can be proved that for this EOS $\rho_c=\rho_c(b)$, $p_c\propto a$ and $T_c \propto a$. Due to these relations, in order to make the pressure field variation proportional to $\Delta x^2$, when changing the mesh parameter $\Delta x$ we choose to fix the value of $b$ and to vary $a$ as:
\begin{equation}
a(\Delta x)=a_\text{base}\,\frac{\Delta x^2}{\Delta x_0^2}~,\label{eq:a-deltax}
\end{equation}
where $a_\text{base}$ is a reference value such that $a(\Delta x_0)=a_\text{base}$ at the coarsest mesh ($\Delta x_0$) set for each numerical test. In Table \ref{tab:fields-scaling} we summarize the scaling of each field of interest with $\Delta x$ (or, equivalently, with $\Delta \hat{x}$).

\begin{table}[h]
\centering
\begin{tabular}{llll}
   Symbol & Description & Conversion factor & Field Scaling (lu) \\
    \midrule
     & Length & $C_\ell=\hat{\ell}_0/\ell_0$ & $\mathcal{O}\left(\Delta x^{-1}\right)$\\
     & Time & $C_t=C_\ell\sqrt{g\ell_0}/\hat{u}_0$ & $\mathcal{O}\left(\Delta x^{-2}\right)$\\
    $\rho$ & Density & $C_\rho=\hat{\rho}_c/\rho_c$ & $\mathcal{O}\left(1\right)$\\
    $p$ & Pressure & $C_p=\hat{p}_c/p_c$ & $\mathcal{O}\left(\Delta x^{2}\right)$ \\
    $T$ & Temperature & $C_T=\hat{T}_c/T_c$ & $\mathcal{O}\left(\Delta x^{2}\right)$ \\
    $\lambda$ & Thermal conductivity & $\left(C_\rho C_\ell^4\right)/\left(C_T C_t^3\right)$
    & $\mathcal{O}\left(1\right)$\\
    $C_v$ & Specific heat at constant volume & $C_\ell^2/\left(C_t^2\,C_T\right)$ & $\mathcal{O}\left(1\right)$ \\
    $\eta$ & Viscosity & $C_\ell^2/C_t$ & $\mathcal{O}\left(1\right)$ \\
    $\mathbf{g}$ & Gravitational acceleration & $C_\ell/C_t^2$ & $\mathcal{O}\left(\Delta x^{3}\right)$\\
   {\color{\newtextcolor} $\mathbf{F}^{b}$ }& {\color{\newtextcolor}Body Forces} & {\color{\newtextcolor}$(C_\rho C_\ell)/C_t^2$} & {\color{\newtextcolor}$\mathcal{O}\left(\Delta x^{3}\right)$}
     
\end{tabular}\caption{Scaling of the fields when changing the refinement parameter.}\label{tab:fields-scaling}
\end{table}

In summary, hydrodynamic and thermodynamic similarities imply in 
varying three parameters in order to keep the physical consistency: $a \propto \Delta \hat{x}^2$, $\gamma = \gamma(\Delta \hat{x})$ and $\kappa = \kappa( \Delta \hat{x} ) $. Although the relation of $a$ parameter to grid refinement is clear, $\sigma$ and $\kappa$ parameters dependence on $\Delta x$ is not trivial, and for that reason, they should be computed through numerical procedures.

\subsection{Mesh refinement example}

For a better exposition of the employed methodology, we take as an example the solution of a generic one-dimensional problem using a coarse base grid of $L = 128$ lattices, setting $a = 6.12 \times 10^{-2}$ and $T_r = 0.86\,T_c$. Also, assume one is interested in a fixed mesh ratio equal to $1.5$ for three finer meshes. This results in meshes with $L = 192$, $288$ and 
$432$. Let us remark that these meshes were employed for the Stefan-problem study in Sec.~\ref{sec:numerics-droplet-cooling}.

The first step in the grid refinement procedure is to adjust the Peng-Robinson parameter $a$ for each mesh. This parameter is computed through Eq. \eqref{eq:a-deltax}, which results in 
values of $a$ equal to $6.12 \times 10^{-2}$, $2.72 \times 10^{-2}$, $1.21 \times 10^{-2}$ and $5.37 \times 10^{-3}$
for, respectively, $L = 128$, $192$, $288$ and $432$. This parameter influences the hydrodynamic similarity through the Laplace equation, and also acts on the thermodynamic consistency \citep{LycettBrown_2015,Czelusniak2020}, specially on the vapor branch. 

The thermodynamic consistency may be ensured by solving the planar coexistence curve equation, given by the following expression \citep{Li2013, Li2013b}:
\begin{equation}
    \int_{ \rho_{g} }^{ \rho_{l} } \left( p_{sat} - p_{ \text{\scriptsize EOS}} \right)	\frac{\dot{\psi}}{\psi^{1-16G\sigma}} d \rho = 0~,
    \label{eq:integral_coexistence_curve}
\end{equation}
where, for a given value of $a$, the parameter $\sigma$ \citep{Li2013} is adjusted for this equation to be satisfied. This is the second step of the proposed refinement procedure.

\begin{figure}
    \centering
    \begin{subfigure}[b]{0.45\textwidth}
        \centering
        \includegraphics[width=\textwidth]{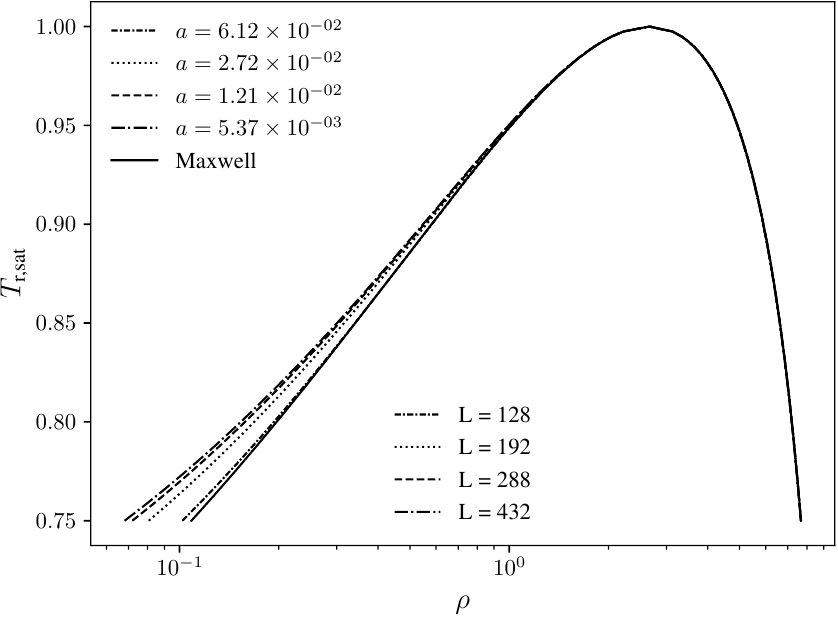}
        \caption{}\label{fig:coexistence-curves-relation-to-a-parameter-a}
    \end{subfigure}
    \hfill
    \begin{subfigure}[b]{0.45\textwidth}
        \centering
        \includegraphics[width=\textwidth]{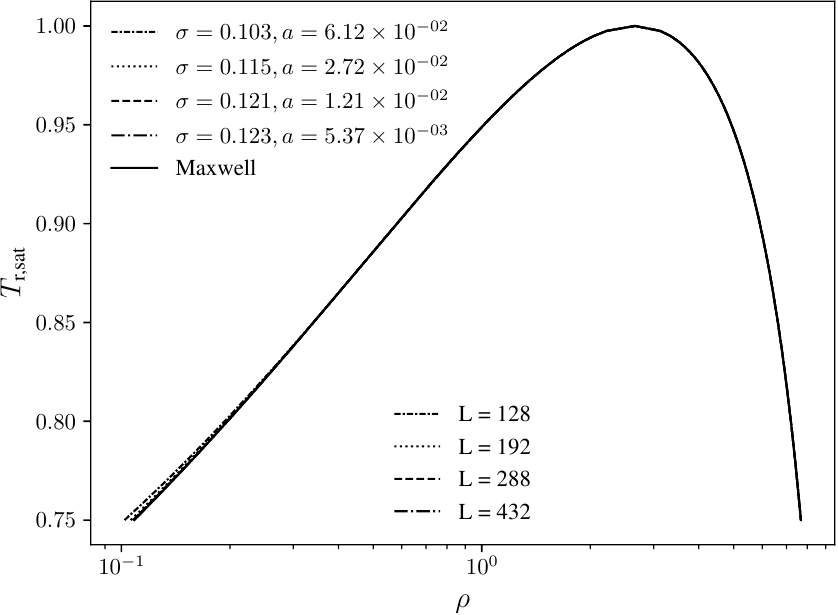}
        \caption{}\label{}
        \label{fig:coexistence-curves-relation-to-a-parameter-b}
    \end{subfigure}
       \caption{\centering Planar interface coexistence curves for distinct Peng-Robinson $a$ parameter values with fixed $\sigma = 0.103$ parameter (a)
       and adjusting $\sigma$ parameter for keeping thermodynamic consistency (b).}
       \label{fig:coexistence-curves-relation-to-a-parameter}
\end{figure}

In Fig.~\ref{fig:coexistence-curves-relation-to-a-parameter-a}, 
coexistence curves for meshes from $L = 128$ to $L= 432$ and their respective values of $a$ are exhibited for $\sigma$ equal to $0.103$, which is a well-known value that ensures a better thermodynamic consistency \citep{Li2013, Li2013b}. In this case, it is possible to see how planar coexistence vapor densities deviate from the Maxwell rule values \citep{clerk1875dynamical}, as Peng-Robinson $a$ parameter decreases. To counteract this effect, $\sigma$ is adjusted for each considered grid. The correct value is found by employing the secant root method to solve Eq. \eqref{eq:integral_coexistence_curve} with an initial guess of $-16\,G\sigma_0 = 2$ \citep{LycettBrown_2015}. Coexistence curves obtained are shown in Fig.~\ref{fig:coexistence-curves-relation-to-a-parameter-b}, where a better agreement between the simulated coexistence curves and the Maxwell rule values is observed, which thus warranties the thermodynamic consistency for each refinement.

The third and last step of the grid refinement procedure consists in ensuring a linear decrease in the surface tension 
$\gamma = \mathcal{O} (\Delta \hat{x})$. For the \cite{Li2013b} surface tension calculation method, defined by Eqs. \eqref{eq:surface-tension-correction-term-C}-\eqref{eq:surface-tension-correction-term-Q}, 
it is possible to show that the surface tension for the planar interface problem is given by:
\begin{equation}
    \gamma = \frac{G c^4}{6} (1-\kappa)\int \psi \frac{d^2 \psi}{dx^2}   \, dx~.
    \label{eq:surface_tension}
\end{equation}

\begin{figure}[h]
	\centering
	\includegraphics[width=0.6\textwidth]{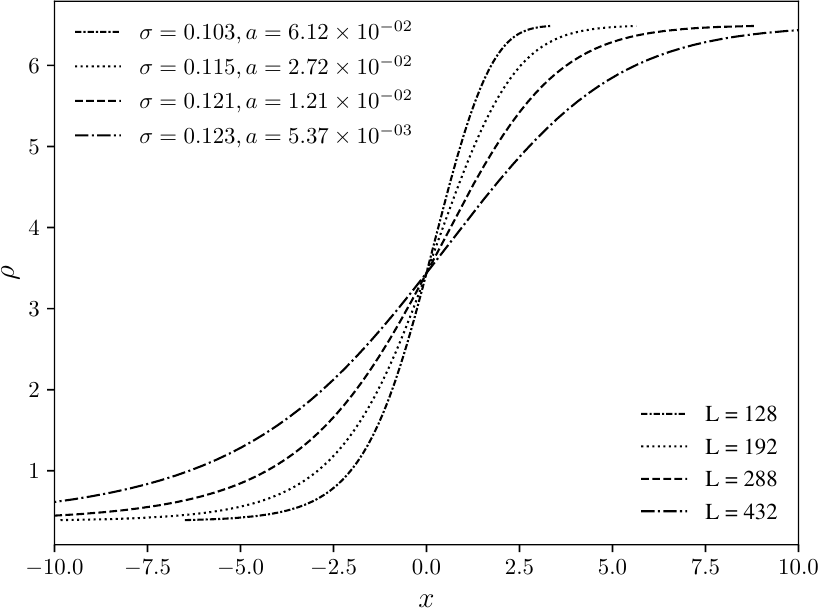}
	\caption{Planar interface density profile for Peng-Robinson EOS with $a$ parameter 
    equal to $6.12 \times 10^{-2}$, $2.72 \times 10^{-2}$, $1.21 \times 10^{-2}$ 
    and $5.37 \times 10^{-3}$}.\label{fig:planar-interface-profiles-variation-with-a}
\end{figure}

From Eq. \eqref{eq:surface_tension}, it is possible to verify that the surface tension value for the planar interface problem is dependent on $\psi$ profile across vapor-liquid interface, and consequently, on the density profile as well. In Fig.~\ref{fig:planar-interface-profiles-variation-with-a}, one can verify how the parameters $\sigma$ and $a$ for $L = 128$ to $L = 432$ influence the density profile. As aforementioned, the parameter $\sigma$ ensures same vapor and liquid densities, while EOS $a$ parameter ensures $\Delta P_{\Delta \hat{x}}=\mathcal{O}\left(\Delta\hat{x}^{2}\right)$. However, the variation of these parameters result in a thicker interface in lattice units, so that the surface tension $\gamma$ is influenced as well. So, the procedure to ensure a linear surface tension decrease with $(\Delta \hat{x})$ consists in finding a $\kappa$ parameter so that:

\begin{equation}
  \gamma_{\text{1}} (\kappa) = \gamma_{\text{0}} \frac{\Delta x}{\Delta x_{\text{0}}}~,
  \label{eq:gamma-meshes-relation}
\end{equation}

\noindent where subscripts $0$ and $1$ refer to, respectively, a fine and a coarse mesh. 
The target finer mesh value $\gamma_1$ is computed from Eq. \eqref{eq:gamma-meshes-relation}, 
and then the parameter $\kappa$ is determined from Eq. \eqref{eq:surface_tension} by substituting $\gamma=\gamma_1$ and solving the right hand side integral for the planar interface problem \citep{Li2013,Czelusniak2020}. This procedure results in $\kappa$ equal to 0, $-5.17 \times 10^{-2}$, 
$-7.44 \times 10^{-2}$ and $-8.45 \times 10^{-2}$ for, $L = 128$, $192$, $288$ and $432$, respectively.

\section{Numerical results}
\label{sec:numerical-results}


Liquid/vapor phase-change problems involve many physical phenomena including: hydrodynamic effects, heat transport (convection and diffusion) between phases and with external medium, thermodynamic properties influence on phase-change, surface tension action of liquid/vapor and liquid/vapor-solid interfaces, solid-fluid interaction and contact angle influence, among others. In this section the proposed mesh refinement procedure, presented in Sec. \ref{sec:problem-definition} is applied for simulating three problems related with liquid/vapor phase-change phenomena with heat transfer.

First, two simpler problems are analyzed in order to develop some intuition about the application of the developed mesh refinement procedure. In Sec. \ref{sec:numerics-droplet-cooling} a two-dimensional droplet vaporization in superheated vapor is presented. The droplet is initially put into contact with a superheated vapor resulting in its vaporization due to the heat transfer in the liquid-vapor interface. Then in Sec. \ref{sec:numerics-1d-boiling} a one-dimensional Stefan problem, mimicking a one-dimensional boiling-like problem, is solved for a configuration similar to the Stefan problem presented in \citet{Safari2013}. Finally, in Sec. \ref{sec:numerics-pool-boiling} the pool boiling problem is assessed considering two models for the contact angle. In the solution of each problem the procedure explained in Sec. \ref{sec:physics-meshes} is employed for refining the mesh and the resulting physical parameters (in l.u.) are reported for each analysis. Unless otherwise stated, the results are re-scaled to the coarsest mesh in order to perform the comparisons.

In the Appendix A, a brief description of the computational tool used for obtaining all the numerical results of this work is given. Also, the characteristics of the supercomputers employed are provided together with the execution times of some pool boiling simulations when varying the mesh refinement.



\subsection{Droplet vaporization}
\label{sec:numerics-droplet-cooling}
\begin{figure}[h]
	\centering
	\includegraphics[width=0.5\textwidth]{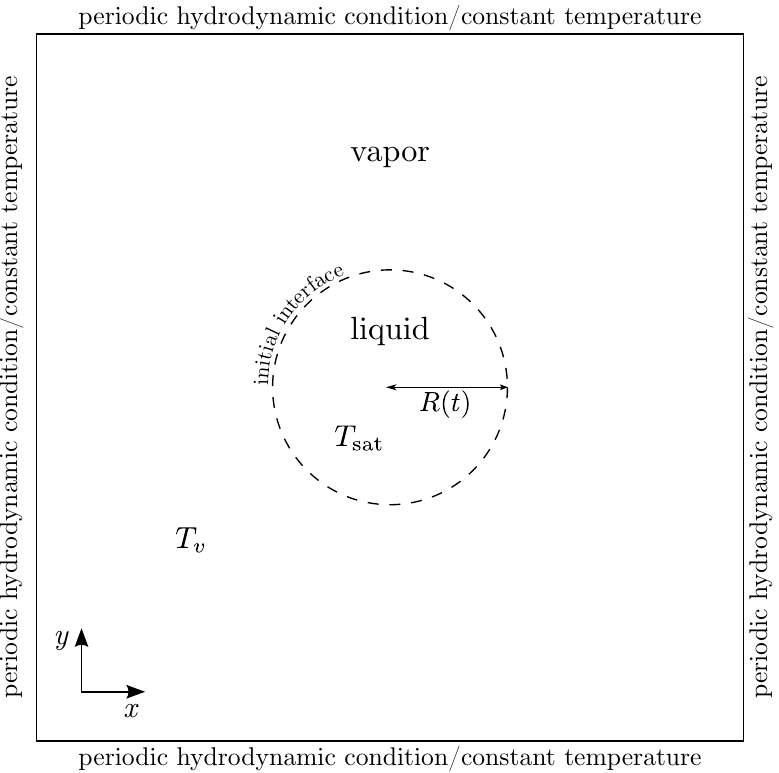}
	\caption{Scheme of the 2D droplet vaporization problem.}\label{fig:droplet-cooling-scheme}
\end{figure}

This problem, schematized in Fig. \ref{fig:droplet-cooling-scheme}, is related to several engineering applications \citep{Fei2022}. Initially, the liquid phase temperature is set to $T_\text{sat}=0.86\,T_c$, while the vapor temperature is set to $T_v=T_c$. The hydrodynamics setting is periodic on both axis while the temperature at the boundary is kept constant and equal to $T_v$. Due to the warmer condition of the vapor (superheated at the same pressure), the liquid evaporates with time and the droplet radius diminishes up-to the collapse.

The droplet is initialized at the center of the domain. At $t=0$, a Heaviside-like function is used for both the temperature and density fields, i.e., for an arbitrary field $\theta$, its shape at $t=0$ is written (where the sub-indices indicate the respective phase)
\begin{equation}
    \theta(\mathbf{x}) = \frac{1}{2}\left\{ \theta_v+\theta_\ell +\left(\theta_\ell-\theta_v\right)\tanh\left(\frac{4.6}{5.0}(d(\mathbf{x},\mathbf{x}_c)-R_0)\right) \right\}~,
\end{equation}
where $d(\mathbf{x},\mathbf{x}_c)$ is the Cartesian distance from $\mathbf{x}$ to the center of the domain ($\mathbf{x}_c$) and $R_0\equiv R(t=0)$ is the initial radius.

For these simulations we fixed the relaxation parameters to $\tau_\rho^{-1}=\tau_e^{-1}=\tau_\zeta^{-1}=\tau_j^{-1}=\tau_q^{-1}=1.25$.

\subsubsection{Grid refinement}

\begin{figure}[h]
	\centering
	\includegraphics[width=0.6\textwidth]{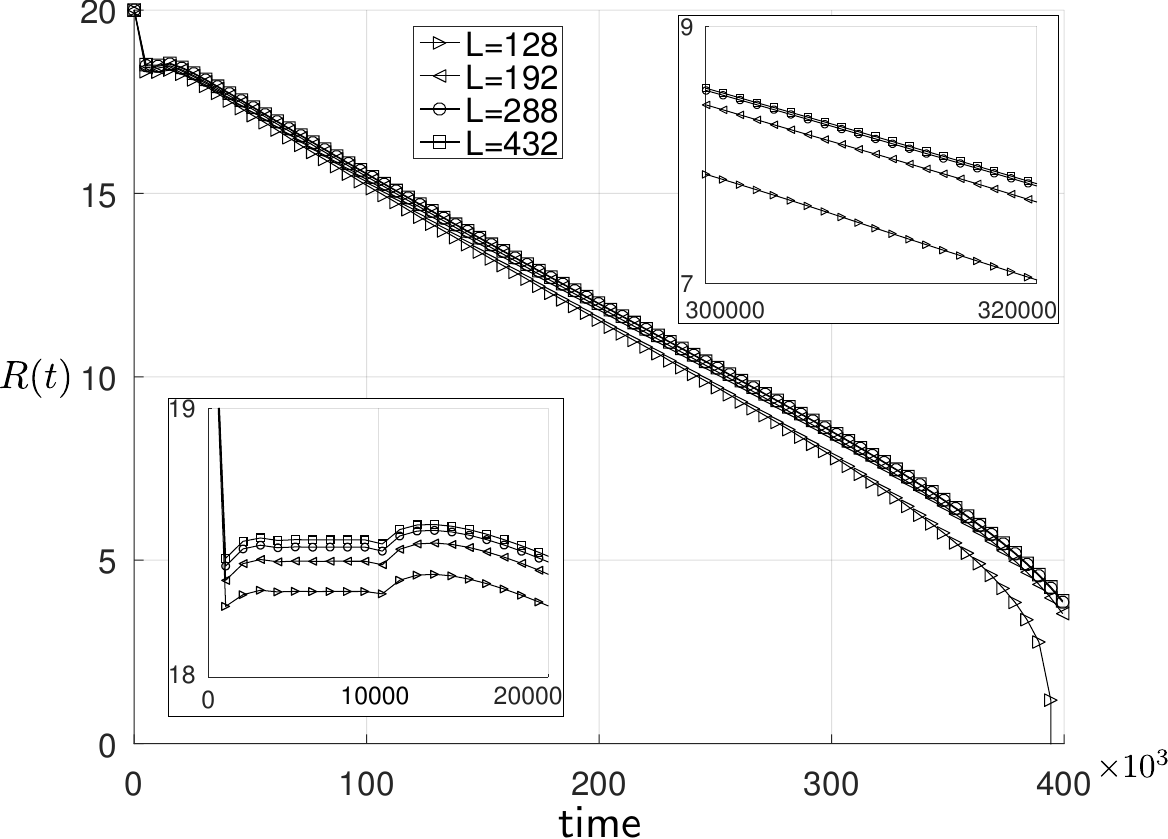}
	\caption{Evolution of the droplet radius in time, $R(t)$, for the four different meshes.}\label{fig:droplet-cooling-meshes}
\end{figure}

The results obtained for four different meshes are reported next. In Fig. \ref{fig:droplet-cooling-meshes}, it is shown how the radius of the droplet $R(t)$ changes with time, while the respective parameters in l.u. are given in Table \ref{tab:parameters-droplet-cooling}. The following parameters were assumed constant for all meshes: $\eta_\ell=\eta_v=0.1$, $\lambda_\ell=\lambda_v=2/3$, $C_{\text{V};\ell}=C_{\text{V};v}=5.0$ and $\mathbf{g}=0$, {\color{\newtextcolor}for these cases, the Jakob number is $\text{Ja}=C_{\text{v}}\left(T_v-T_\text{sat}\right)/h_{fg}=0.133$}. To assure {\color{\newtextcolor} the stability of the droplet} at the beginning of the simulation, the energy equation is not solved ({\color{\newtextcolor}isothermal condition}) from $t=0$ to $t=10.000$. Thus, the temperature field is constant along that time interval. It can be observed that $R(t)$ first oscillates and the mechanical equilibrium is reached afterward at $t\approx 3.000$.

 
\begin{table}[h]
\centering
\begin{tabular}{c|cccc|c}
\toprule
L& steps $\times 10^6$ & $a$ & $\sigma$ & $\kappa$ & $\Delta^* R(t)$ \\
\midrule 
128&  0.40 & $6.12\times 10^{-3}$ & 0.103 & 0  
& - \\
192&  0.90 & $2.72\times 10^{-3}$ & 0.115 & $-5.17\times 10^{-2}$ & $2.83\times 10^{-2}$ \\
288&  2.03 & $1.21\times 10^{-3}$ & 0.121 & $-7.44\times 10^{-2}$ & $6.64\times 10^{-3}$ \\
432&  4.56 & $5.37\times 10^{-4}$ &  0.123 & $-8.45\times 10^{-2}$ & $1.35\times 10^{-3}$ \\
\bottomrule
\end{tabular}\caption{Lattice units parameters for the refinement of the droplet vaporization.}
\label{tab:parameters-droplet-cooling}
\end{table}

The curve $R(t)$ displayed in Fig. \ref{fig:droplet-cooling-meshes} show a fast convergence as a function of mesh size. This behavior is confirmed by the data presented in Table \ref{tab:parameters-droplet-cooling} for a relative quantity $\Delta^* R(t)$, defined as:

\begin{equation}
\Delta^* R(t)= \frac{\sum_{t=0}^T |R(t)-R^*(t)|}{\sum_{t=0}^T |R^*(t)|}~,
\label{eq:relative-change-R}
\end{equation}

In Eq. \eqref{eq:relative-change-R} relation $R(t)$ is evaluated at a certain mesh and $R^*(t)$ is evaluated at its immediate refined mesh. The values of $\Delta^* R(t)$ are shown in the last column of Table \ref{tab:parameters-droplet-cooling}. The value of the second row ($2.83\times 10^{-2}$) corresponds to the relative difference between the curves for $L=128$ and $L=192$, while the value at the third row corresponds to the difference between $L=192$ and $L=288$, and that presented in the last row to the difference between $L=288$ and $L=432$, respectively. It can be observed that the results present an asymptotic regime of convergence (related to the different meshes) as the relative difference decreases at a ratio approximately constant. Indeed:
$$\frac{\Delta^* R(t)_{128,192}}{\Delta^* R(t)_{192,288}}=\frac{2.83\times 10^{-2}}{6.64\times 10^{-3}}\approx 4.27\,,\qquad \frac{\Delta^* R(t)_{192,288}}{\Delta^* R(t)_{288,432}}=\frac{6.64\times 10^{-3}}{1.35\times 10^{-3}}\approx 4.91~,$$
which gives convergence orders of 
$$\log(4.27)/\log(1.5)\approx 3.58,\qquad \log(4.91)/\log(1.5)\approx 3.92~,$$
respectively. Let us recall that this convergence order is related to both spatial and time refinement as the viscous regime is being imposed, where $\Delta t\propto \Delta x^2$.

\begin {comment}
For this problem, a fast convergence is observed for the curve $R(t)$. In fact, defining $\Delta^* R(t)$ as the relative quantity
\begin{equation}
\Delta^* R(t)= \frac{\sum_{t=0}^T |R(t)-R^*(t)|}{\sum_{t=0}^T |R^*(t)|}~,
\label{eq:relative-change-R}
\end{equation}
where $R(t)$ is evaluated at a certain mesh and $R^*(t)$ is its immediate refinement, $\Delta^* R(t)$ is shown in the last column of Table \ref{tab:parameters-droplet-cooling}. The value of the second row ($2.83\times 10^{-2}$) corresponds to the relative difference between the curves for $L=128$ and $L=192$; the value at the third row corresponds to the difference between $L=192$ and $L=288$, and so on. One can observe that the results seem to be at an asymptotic regime (related to the different meshes) as that relative difference decreases at a ratio approximately constant, in fact:
$$\frac{\Delta^* R(t)_{128,192}}{\Delta^* R(t)_{192,288}}=\frac{2.83\times 10^{-2}}{6.64\times 10^{-3}}\approx 4.27\,,\qquad \frac{\Delta^* R(t)_{192,288}}{\Delta^* R(t)_{288,432}}=\frac{6.64\times 10^{-3}}{1.35\times 10^{-3}}\approx 4.91~,$$
which gives convergence orders of 
$$\log(4.27)/\log(1.5)\approx 3.58,\qquad \log(4.91)/\log(1.5)\approx 3.92~,$$
respectively. Let us recall that this convergence order is related to both spatial and time refinement as we are imposing the viscous regime, where $\Delta t\propto \Delta x^2$.

This high convergence order was the highest one observed during this work. Also, it was the only case where phase-change and, at the same time, evidence of an asymptotic convergence regime were found. It is worth noticing that on this configuration no effect related to wall nodes (where fluid-solid interaction takes place) is involved.
\end {comment}


\subsection{One-dimensional boiling-like}
\label{sec:numerics-1d-boiling}

\begin{figure}[h]
	\centering
	\includegraphics[width=0.5\textwidth]{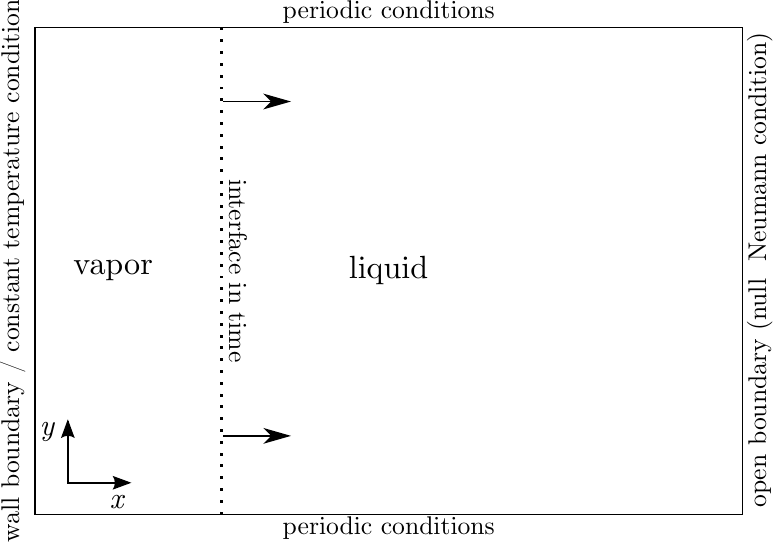}
	\caption{Scheme of the one-dimensional boiling-like problem.}\label{fig:stefan-problem-scheme}
\end{figure}

The second problem analyzed consists of a one-dimensional configuration where the fluid liquid phase is initially in equilibrium at the saturation temperature $T_\text{sat}=0.86\,T_c$ on the whole computational domain. For $t> 0$, the temperature at the left wall ($T_w$) assumes a value higher than $T_\text{sat}$ and a phase-change from liquid to gas is triggered forming a liquid/vapor interface as shown in Fig. \ref{fig:stefan-problem-scheme}. This phase-change process produces the interface movement from left to right, reaching the right end of the domain at a finite time, where null Neumann boundary conditions \citep{Lou2013} are imposed.


In the idealized incompressible limit, the temperature on the liquid phase region is homogeneous and constant in time, being equal to the initial saturation temperature of the fluid $T_\ell\equiv T_\text{sat}$, while in the vapor region the temperature decays linearly from $T_w$ to the temperature of the liquid. This is the so-called one-dimensional Stefan problem, see for instance \citet{Safari2013}. The solution to this problem is obtained by solving the energy conservation equation for each phase and applying the boundary conditions at domain boundaries and the interface \citep{Alexiades1993}. This last condition assumes a continuity of both mass and energy fluxes at the interface. The analytical solution results in the following expression for the interface position in time:
\begin{equation}
s(t) = 2\beta \sqrt{\alpha_v \,t}~,
\label{eq:stefan-analytical}
\end{equation}
where $\alpha_{\theta}\equiv \lambda_{\theta}/(c_{\text{v},\theta}\,\rho_{\theta})$, (for $\theta=\ell,v$) denotes the thermal diffusivity of each phase and $\beta$ is the solution of the following equation:
\begin{equation}
    \beta \exp(\beta^2)\,\text{erf}(\beta)=\frac{c_{\text{v},v}(T_w-T_\ell)}{h_{fg}\sqrt{\pi}}~,
\end{equation}
being $c_{\text{v},v}$ the specific heat at constant volume of the vapor and $h_{fg}$ the latent heat of vaporization. This last property is computed from the Peng-Robinson EOS following \cite{Gong2013} .

\subsubsection{Grid refinement and compressibility effects}
Two set of parameters were tested: For the first set, the physical parameters are fixed as in Section \ref{sec:numerics-droplet-cooling} excepting the thermal conductivities, that are set such that the thermal diffusivities correspond to $\alpha_\ell=\alpha_v=0.3$, and the wall temperature is set to $T_w=0.9\,T_c$. For the second set of parameters, it was imposed that $c_{\text{v},\ell}=c_{\text{v},v}=10$, $\lambda_\ell=15$ and $\lambda_v=1$ in order to keep the diffusivities relatively constant ($\alpha_\ell=0.23$ and $\alpha_v=0.26$) while diminishing the effect of the compression term in Eq. \eqref{eq:energy-equation}. {\color{\newtextcolor} The Jakob number is $\text{Ja}=0.038$ for the first set of parameters and $\text{Ja}=0.076$ for the second one.} On the left side, using the $\psi$-based contact angle model (see Sec. \ref{sec:mathematical-modeling-contact-angle}), the wall is modeled as being hydrophobic by setting $\rho_w$ to the vapor density $\rho_v$. For these simulations the relaxation parameters are fixed to $\tau_\rho^{-1}=1.0$, $\tau_e^{-1}=\tau_\zeta^{-1}=0.8$, $\tau_j^{-1}=1.0$ and $\tau_q^{-1}=1.1$.

\begin{figure}[h]
	\centering
	\includegraphics[width=0.95\textwidth]{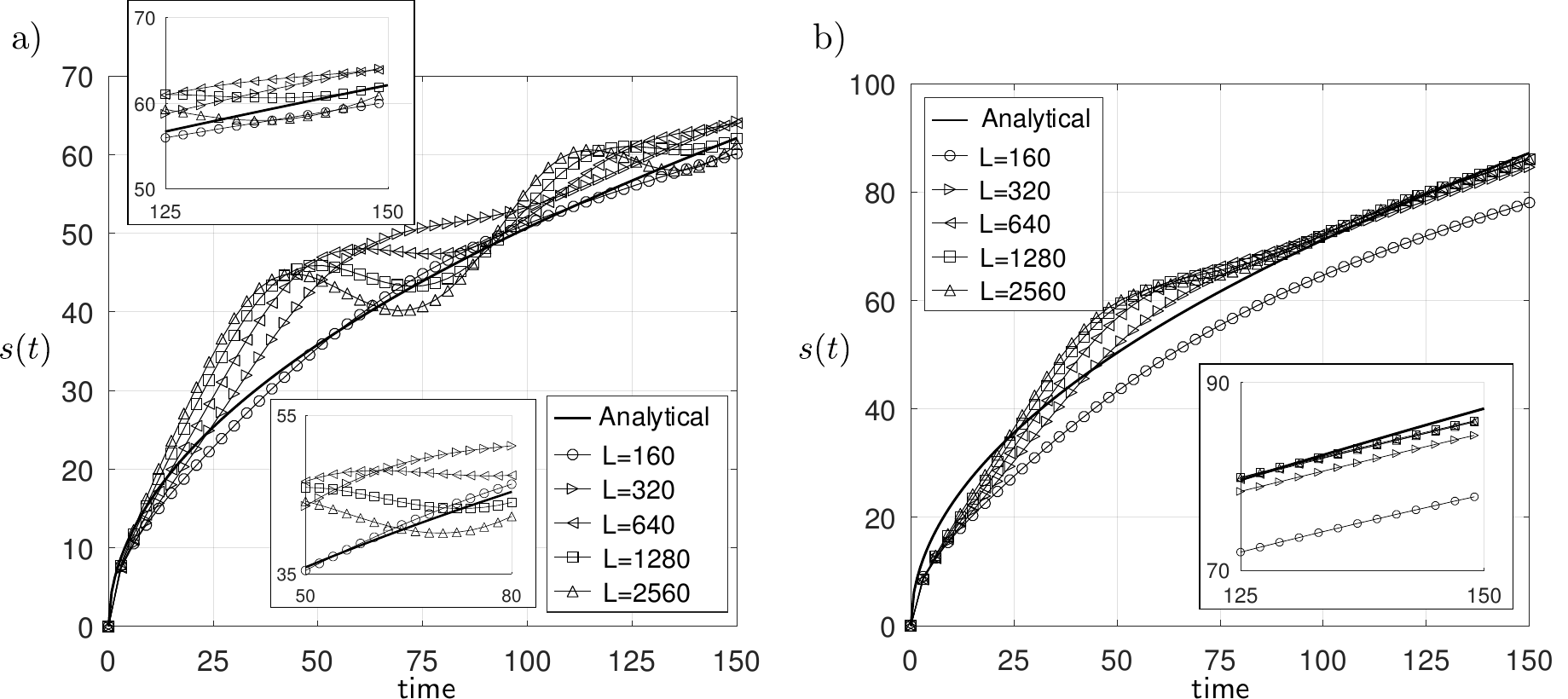}
	\caption{Interface position for the one-dimensional boiling-like problem for different meshes. a) First set of parameters, b) Second set of parameters.}\label{fig:stefan-refinement}
\end{figure}

The resulting curves $s(t)$ considering the first set of parameters are shown in Fig. \ref{fig:stefan-refinement} a) for $L$ varying from 160 to 2560. The analytical solution (Eq. \eqref{eq:stefan-analytical}) is also included. Let us remark that this analytical solution is used here as a reference and, as the energy equation (Eq. \eqref{eq:energy-equation}) incorporates compressibility effects, the numerical solutions obtained by the LBM do not need to converge to this reference. In fact, one can observe that as the number of lattices augments there appear oscillations of $s(t)$ around the reference solution. Moreover, for this first set of parameters, there exist some time intervals where the interface speed $s'(t)$ is negative.

The relative change between the curves is computed analogously to Eq. \eqref{eq:relative-change-R} and is shown in Table \ref{tab:relative-change-stefan} (row 1st set of parameters). It can be observed that the convergence ratio for this set of parameters is slow. Moreover, the convergence order varies between 0.89 for the coarsest meshes and 0.24 for the finest ones, which indicates that these results do not correspond to an asymptotic convergence regime (which requires the convergence order to be invariant among meshes \citep{Debortoli2015}). 

The oscillations of the liquid/vapor interface seem to be a consequence of the fluid compressibility. Numerical evidence supporting this assertion was first obtained by using the second set of parameters, for which the effects of the compression term in Eq. \eqref{eq:energy-equation} are reduced. The relative change between the curves when refining the mesh for this second set of parameters is shown in Table \ref{tab:relative-change-stefan} (row 2nd set of parameters) and the corresponding curves $s(t)$ are shown in Fig. \ref{fig:stefan-refinement} b). This time the oscillations are less pronounced and a faster convergence is observed, with a convergence order that goes from 3.28 for the coarsest meshes to 1.59 for the finest ones. These results are a sign of these interface position oscillations being not numerical artifacts but a consequence of the physics constituting the LBM simulation model used in this work.

\begin{table}[h]
\centering
\begin{tabular}{c|cccc}
\toprule
 &  320 & 640 & 1280 & 2560 \\
\midrule
$\Delta^*_1 s(t)$ (1st set) & $8.86\times 10^{-2}$ & $4.77\times 10^{-2}$ & $4.57\times 10^{-2}$ & $3.89\times 10^{-2}$ \\
$\Delta^*_2 s(t)$ (2nd set) & $1.09\times 10^{-1}$ & $3.31\times 10^{-2}$ & $1.47\times 10^{-2}$ & $9.28\times 10^{-3}$ \\
\bottomrule
\end{tabular}\caption{Relative change of the curves $s(t)$ between meshes (from $L=320$ to $L=2560$).}
\label{tab:relative-change-stefan}
\end{table}

\begin{figure}[t]
	\centering
	\includegraphics[width=0.95\textwidth]{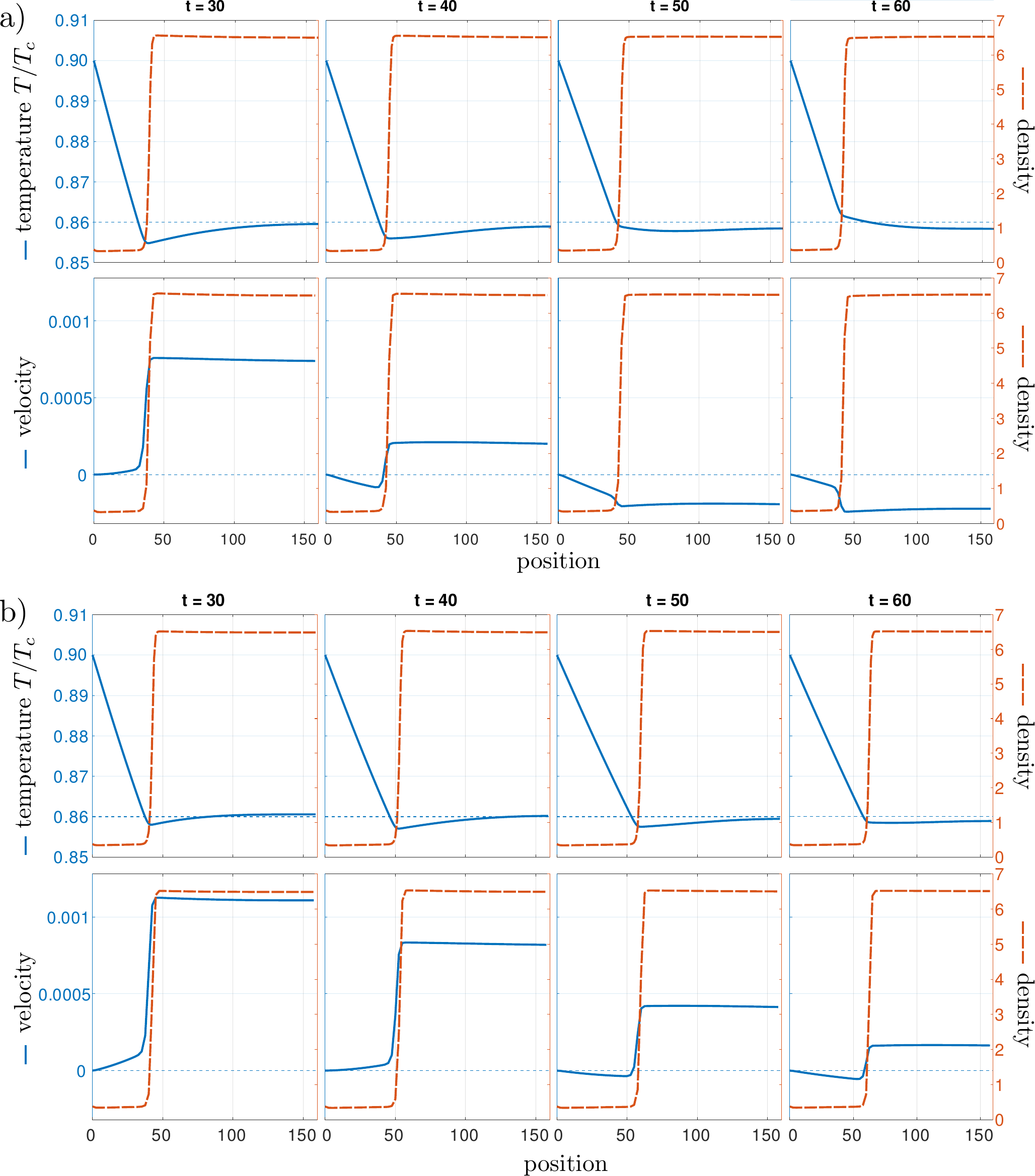}
	\caption{Interface position for the one-dimensional boiling-like problem for different meshes. a) First set of parameters, b) Second set of parameters.}\label{fig:stefan-profiles}
\end{figure}

Following a procedure similar to the one exposed by \cite{Welch2000}, one can find that continuity of the mass and energy fluxes, at the idealized incompressible limit, imply the following equation for the interface velocity, ${ds}/{dt}$:

\begin{equation}
    h_{fg}\rho_v \frac{ds}{dt}= -\lambda_v\left. \parder{T}{x}\right|_{x=s^-}+h_{fg}\rho_v\left.u\right|_{x=s^-}+\lambda_\ell\left. \parder{T}{x}\right|_{x=s^+}~,\label{eq:stefan-analytic-complete}
\end{equation}
where the superscripts ``$+,-$" indicate the right or left limits respectively and $u$ denotes the fluid velocity (liquid or vapor phases). The last two terms of this equation are null at the incompressible limit. In fact, disregarding these terms and integrating the resulting equation one recovers the formula \eqref{eq:stefan-analytical} \citep{Welch2000}. Nevertheless, Eq. \eqref{eq:stefan-analytic-complete} can be used to obtain some insight in our LBM simulations where compressible effects are not negligible.

The density, temperature, and velocity profiles of the fluid, for the first set of parameters and a series of timesteps, are shown in Fig. \ref{fig:stefan-profiles} a) around the instant of time where the first interface oscillation takes place. Contrary to what happens in the incompressible limit, the interface temperature oscillates in time around $T_\text{sat}$. Let us remark from the last term in Eq. \eqref{eq:energy-equation} that the vaporization process tends to cool the interface, while the liquefaction process has the opposite effect. One can notice that this temperature different from $T_\text{sat}$ at the interface produces non-null gradients of both the fields $T$ and $\rho$ (and thus non-null velocities due to pressure gradients) that make the LBM solution differ from the analytical one. For instance, at $t=50$ the derivative $\parder{T}{x}$ at the right side of the interface is positive while the velocity of the fluid at the left side (vapor region) is negative. Considering Eq. \eqref{eq:stefan-analytic-complete} and these two facts, one could expect the velocity of the interface to decrease (relative to the incompressible regime) or even to make it reverse its movement, as observed around that instant of time. This relation between the oscillation of the interface and these non-null gradients, depending on whether $\left.T\right |_{x=s(t)}$ is greater or lower than $T_\text{sat}$ was observed through the whole simulation.

The profiles for $\rho$, $T$, and $u$ for the same timesteps but for the second set of parameters are shown in Fig. \ref{fig:stefan-profiles} b). As for the case discussed above, it was observed that $\left.T\right|_{x=s(t)}$ is lower than $T_\text{sat}$ for the first time instants too, but its value is slightly higher. As a consequence, this time the gradients in $T$ and $\rho$ are lower, and also the vapor's velocity near the interface (in absolute value), which has the overall effect of the LBM solution to be more similar to the analytic solution given by Eq. \eqref{eq:stefan-analytical}.

\subsection{Pool boiling}
\label{sec:numerics-pool-boiling}

In this section, the refinement procedure described in Sec. \ref{sec:problem-definition} is applied for the pool boiling problem. The numerical tests presented were performed using the $\psi$-based contact angle model. To justify this choice, let us first show results of refinement tests considering the physical phenomena related to the wall/fluid interaction alone and the two contact angle models presented in Section \ref{sec:mathematical-modeling}. For the results shown hereafter the timescale will be fixed by always giving the time values relative to the coarsest mesh considered, so the timestep value of a certain fine mesh is re-scaled considering that $\Delta x=\Delta t^2$. For these simulations we fixed the relaxation parameters to $\tau_\rho^{-1}=1.0$, $\tau_e^{-1}=\tau_\zeta^{-1}=0.8$, $\tau_j^{-1}=1.0$ and $\tau_q^{-1}=1.1$.

\subsubsection{Contact angle model results}
\label{sec:numerics-contact-angle}

\begin{figure}[h]
	\centering
	\includegraphics[width=0.9\textwidth]{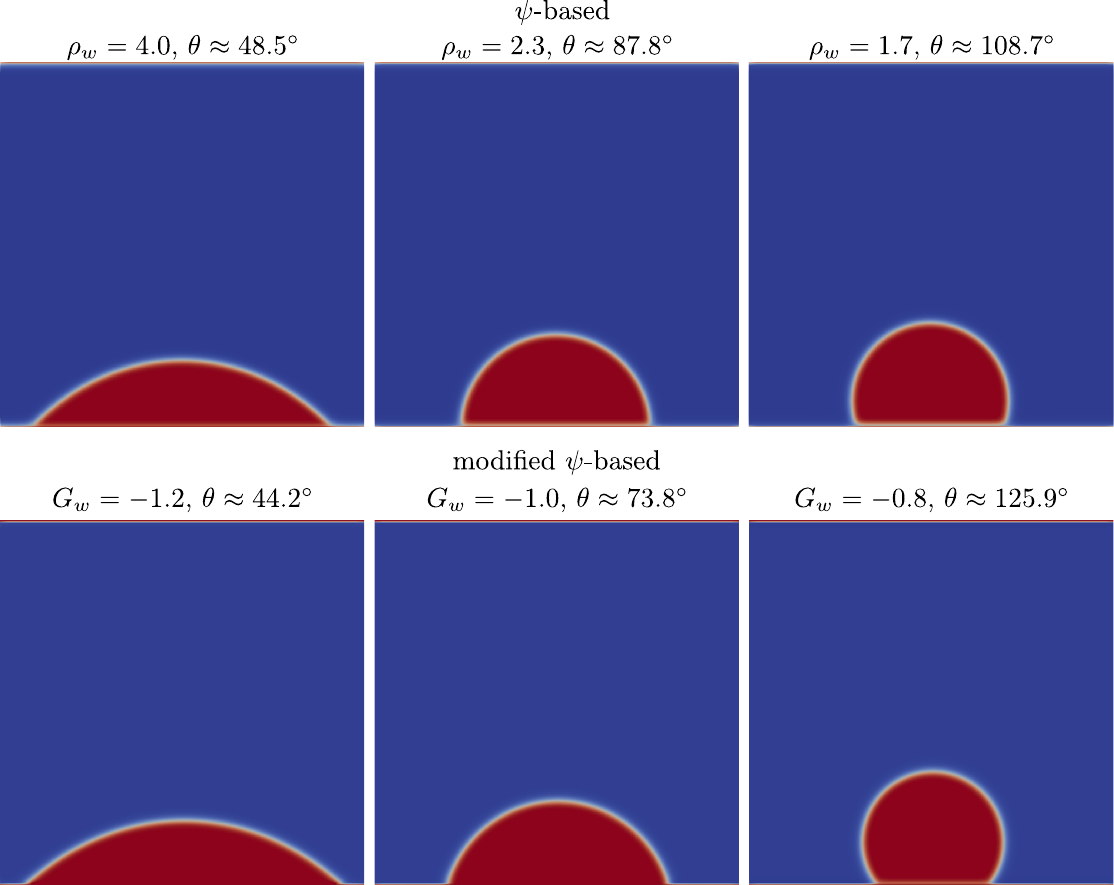}
	\caption{Resulting droplet density fields and contact angles for two contact angle models for different surfaces smoothness.}\label{fig:contact-angle-profiles}
\end{figure}

For these tests, a droplet is initialized near a wall where the fluid/solid interaction takes place. {\color{\newtextcolor}The tests run a fixed number of timesteps large enough for the simulations to reach a static regime. To test that our implementations work as expected, a first set of simulations is run for a mesh of $160\times 160$ varying the different contact angle parameters from hydrophilic to hydrophobic configurations. The resulting density profiles are shown in Fig. \ref{fig:contact-angle-profiles} for both the $\psi$-based and the modified $\psi$-based models and three different contact angles. The results are in line with the expected according to \cite{Li2014}.}

\begin{figure}[h]
	\centering
	\includegraphics[width=1.0\textwidth]{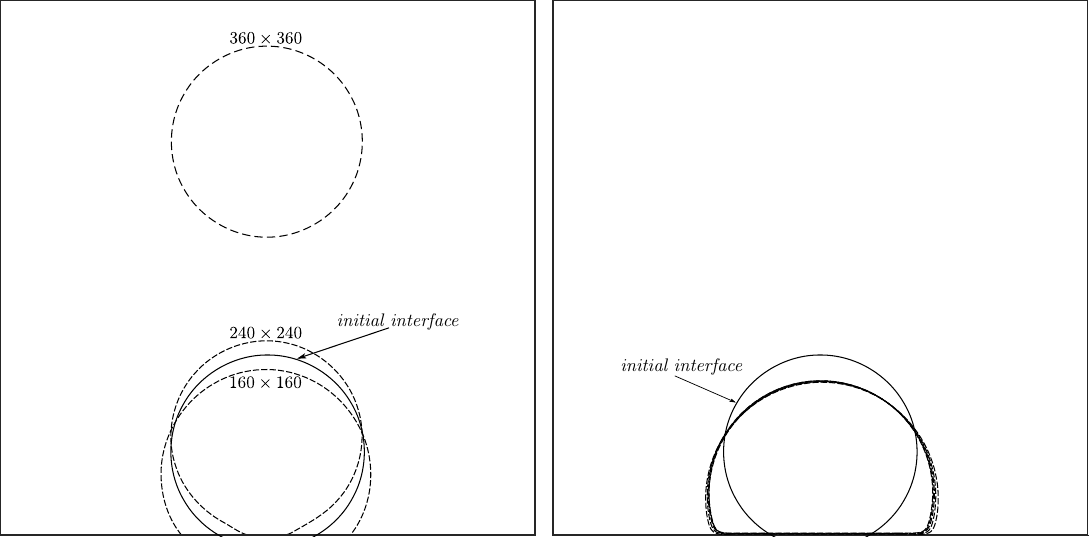}
	\caption{Resulting static liquid/vapor interfaces for different meshes. Left: for the modified $\psi$-based model with $G_w=-0.8$. Right: for the $\psi$-based model with $\rho_w=1.7$. The initial interface is also displayed by a continuous curve.}\label{fig:modified-psi-based-profiles-meshes}
\end{figure}

This initial $160\times160$-lattices mesh was refined four times, using meshes of $240\times 240$, $360\times 360$, $540\times 540$, and $810\times 810$ lattices (corresponding to a refinement ratio of 1.5). The relative change of the density fields (when reaching the static regime) is shown in Table \ref{tab:relative-change-contactangle} for the results obtained along with the $\psi$-based model. These results indicate a convergent behavior of the $\psi$-based contact angle model without the need to adjust its defining parameter $\rho_w$, with a convergence rate of around 0.9. On the other hand, the simulations performed for the modified $\psi$-based model show that the parameter $G_w$ needs to be adjusted to keep the contact angle constant when applying the proposed refinement procedure. This can be observed in Fig. \ref{fig:modified-psi-based-profiles-meshes}  where the static liquid/vapor interface is shown for three different meshes and $G_w=-0.8$ (left), {\color{\newtextcolor}and $\rho_w=1.7$ (right)}. {\color{\newtextcolor}Up to our knowledge, there does not exist a simple analytic or numerical method to determine this parameter for a given mesh. To exemplify this, fixing the contact angle to 125.9$^\circ$, the corresponding values of $G_w$ were explored by employing a linear search for different meshes, the obtained values are shown in Table \ref{tab:gws}. It is possible to observe that the behavior of $G_w$ in terms of the mesh is non-trivial. Notice that this linear search is expensive, as to determine each of these values one needs to run a series of droplet-solid tests. For these reasons, and to simplify the exposition, the following Pool Boiling results are presented by using only the $\psi$-based model.}

\begin{table}[b]
\centering
\begin{tabular}{lcccc}
\toprule
$N$ & 160 & 240 & 360 & 540 \\
$G_w$ & -0.800 & -0.865 & -0.910 & -0.940 \\
\bottomrule
\end{tabular}\caption{Values of $G_w$ fixing the contact angle for different meshes.}\label{tab:gws}
\end{table}

\begin{table}[h]
\centering
\begin{tabular}{c|cccc}
\toprule
 &  240 & 360 & 540  & 810 \\
\midrule
$\Delta^* \rho$ ($\rho_w=4.0$) & $10.4\times 10^{-2}$ & $8.7\times 10^{-2}$ & $6.3\times 10^{-2}$ & $4.4\times 10^{-2}$\\
$\Delta^* \rho$ ($\rho_w=2.3$)& $12.9\times 10^{-2}$ & $10.1\times 10^{-2}$ & $7.1\times 10^{-2}$ & $4.9\times 10^{-2}$\\
$\Delta^* \rho$ ($\rho_w=1.7$) & $12.4\times 10^{-2}$ & $9.1\times 10^{-2}$ & $6.0\times 10^{-2}$  & $4.1\times 10^{-2}$\\
\bottomrule
\end{tabular}\caption{Relative change of the density profiles for the $\psi$-based contact angle model.}
\label{tab:relative-change-contactangle}
\end{table}

\subsubsection{Pool boiling results}
\label{sec:numerics-pool-boiling-results}
\begin{figure}[h]
	\centering
	\includegraphics[width=0.5\textwidth]{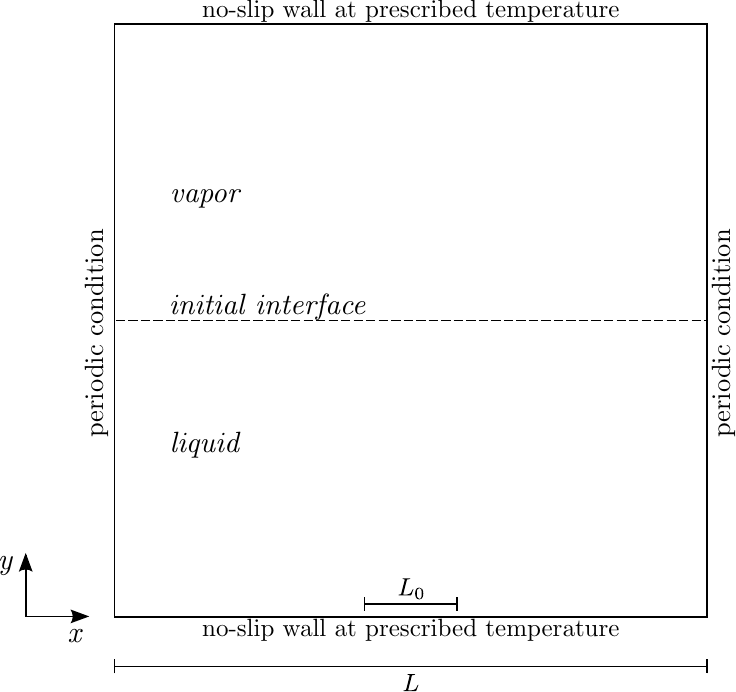}
	\caption{Scheme of the 2D pool boiling problem, boundary and initial conditions for the phases.}\label{fig:pool-boiling-scheme}
\end{figure}

The two-dimensional configuration of the simulated pool boiling problem is schematized in Fig. \ref{fig:pool-boiling-scheme}. Periodic boundary conditions are set at both the left and right boundaries ($\mathbf{x}=0$ and $\mathbf{x}=L$)  and the wet-node schemes of \cite{Zou1997} are used to prescribe no-slip walls at the bottom and top positions. A region of $L_0$ lattices is identified at the middle of the bottom wall to impose a particular condition either for temperature or for the contact angle. 

The initial interface vapor/liquid is set at half the domain height. Thus, the initial conditions at the interior of the domain are:
\begin{equation}
\hat{T}(\mathbf{x},t=0)=\hat{T}_\text{sat}~,\qquad
    \hat{\rho}(\mathbf{x},t=0)=\begin{cases}
    \hat{\rho}_\ell & \text{if}~y<L/2\\
    \hat{\rho}_v & \text{otherwise}
    \end{cases}~,\label{eq:boundary-cond}
\end{equation}
along with a null velocity field.

At the upper wall, the temperature is fixed to the saturation temperature, which is set to $T_\text{sat}=0.86\,T_c$. The same temperature is set at the lower wall, excepting for the heater region (placed at half of the domain and with a length equal to $L_0$) where the temperature, denoted by $T_h$, assumes two possible values $T_h=1.0\,T_c, 1.2\,T_c$, {\color{\newtextcolor}for which $\text{Ja}=0.133$ and $\text{Ja}=0.153$ respectively. The Reynolds number for these simulations is 39.3 for the liquid phase and 2.3 for the vapor phase (to compute these numbers the surface tension is needed, which is determined by means of the static droplet test)}. The subscript $h$ is used to denote the parameter value at the heater. 

    Regarding the contact angle along the surfaces, the upper wall was set as being hydrophobic by imposing $\rho_w=0.38$ there. At the lower surface, for the wall lattices that are outside the heater, the surface was modeled as being hydrophilic by setting $\rho_w=6.50$. At the heater lattices, two configurations were tested: one hydrophobic, where $\rho_w\equiv \rho_{wh}=0.38$, and one \emph{intermediate}, where $\rho_w\equiv \rho_{wh}=3.44$. This gives a total of four possible configurations.
    
    The baseline set of parameters for the next simulations corresponds to $160\times 160$ lattices, for which $a=6.12\times 10^{-3}$, $\sigma=0.103$, $\kappa=0$, $c_{\scriptsize{\text{V}}}=5$, $g=-2.5\times 10^{-5}$, $\eta_\ell=\eta_v=0.1$, $\lambda_\ell=1.95$ and $\lambda_v=0.114$. Remembering that some of these parameters change when refining the mesh according to Section \ref{sec:physics-meshes}.

\subsubsection*{Qualitative comparisons}
\label{sec:numerics-boil-qualitative-comparisons}

\begin{figure}[h]
	\centering
	\includegraphics[width=0.95\textwidth]{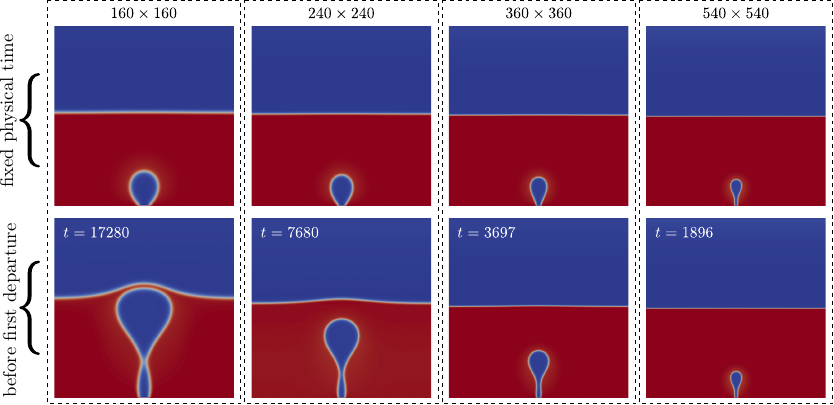}
	\caption{Density profiles obtained by refining a simulation of the pool boiling problem without adjusting the physical parameters among meshes.}\label{fig:pool-without-adjusting}
\end{figure}

Let us first show what happens if one just changes the number of lattices without adjusting of the parameters ($a$, $\sigma$, $\kappa$ and $g$). An example of this is shown in Fig. \ref{fig:pool-without-adjusting} for the case $\rho_{wh}=0.38$ and $T_h=1.2\,T_c$. In that figure, the spatial and temporal meshes are refined from left to right. In the first row of results the physical time is fixed (considering $\Delta t \propto \Delta x^2$) at a time instant where the first bubble is developing; and at the second row of results, the physical time is set for the instant just before the departure of the bubble in each simulation. The time values that are shown in that last row correspond to the time values respective to the coarsest mesh ($160\times 160$) considering $\Delta t \propto \Delta x^2$. From these results, it is clear that the physical behavior is not consistent when refining the mesh, as the bubble shape, size, and time departure have no relation among meshes.

\begin{figure}[h]
	\centering
	\includegraphics[width=0.95\textwidth]{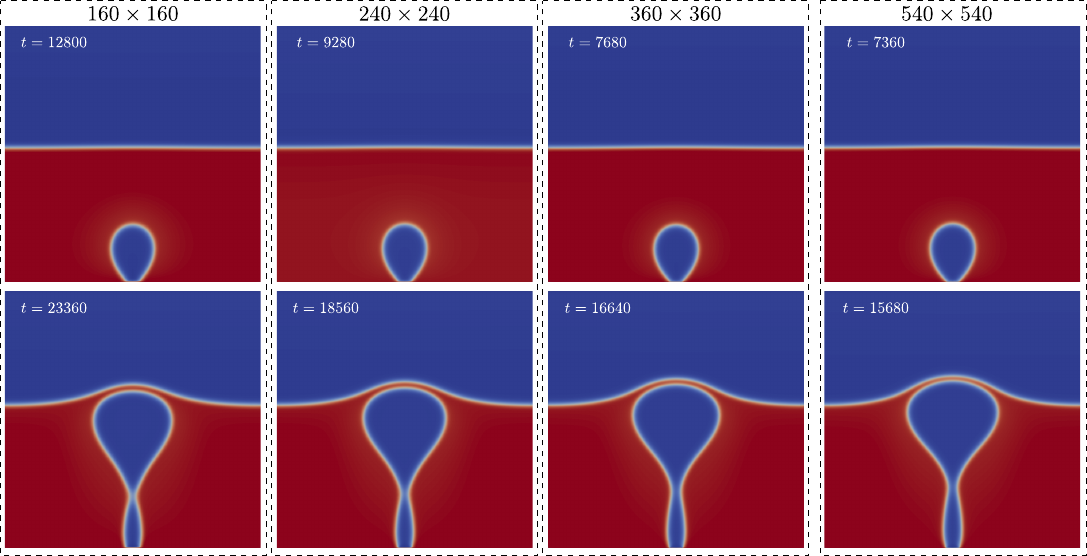}
	\caption{Density profiles obtained by refining a simulation of the pool boiling problem adjusting the physical parameters among meshes according to Section \ref{sec:physics-meshes}.}\label{fig:pool-adjusting}
\end{figure}

Repeating these simulations but this time adjusting the physical parameters as described in Section \ref{sec:physics-meshes} the results show a coherent relation among meshes. This is shown qualitatively in Fig. \ref{fig:pool-adjusting} for $\rho_{wh}=0.38$ and $T_h=1.2\,T_c$, where the first row of snapshots corresponds to an instant of time (all the timestep values are re-scaled to the coarsest mesh) in which the first bubble is developing and its boundary is far from the initial interface, and the second row corresponds to an instant of time just before the bubble's departure takes place. Let us remark two facts that can be observed from these results: 1) for the density profiles obtained for the coarsest mesh at both $t=12800$ and $t=23360$, there exists a timestep at each refined simulation that is similar to the coarsest mesh density profile, and 2) there exists a \emph{delay}) between the simulations that diminishes as $\Delta t$ goes to 0. In fact, denoting by $\delta$ the value of this delay at an arbitrary timestep, for the event of the first row (the growing bubbles) one has $\delta=12800-9280=3520$ between the two first meshes, $\delta=9280-7680=1600$ between the configurations $360\times 360$ and $240\times 240$, and $\delta=7680-7360=320$ between the last pair of finer meshes. A similar behavior can be computed for the time instant where the first bubble is about to depart, this time $\delta$ assumes the values $4800$, 1920, and 960 from the coarser to the finer pair of consecutive grids. These qualitative observations are numerically assessed in the following section.

\subsubsection*{Time and space differences between meshes}
\label{sec:pool-global-errors}

To analyze quantitatively the convergence of the fields of interest let us introduce the space relative difference for an arbitrary field $\theta=\theta(\mathbf{x},t)$ between two meshes:
\begin{equation}
    \text{d}^{N} \theta (t) = \frac{\sum_{i\in \mathcal{S}}\left| \mathcal{I}(\theta^{N})(x_i,t)-\theta^{N^*}(x_i,t)\right|}{\sum_{i\in \mathcal{S}}\left| \theta^{N^*}(x_i,t)\right|}~,\label{eq:rel-diff-timestep}
\end{equation}
for arbitrary $N$ and $N^*$ denoting the immediate finer mesh (e.g., for $N=160$ one has $N^*=240$), and where $\mathcal{S}$ refers to the spatial points indices of the fine mesh with $N^*\times N^*$ lattices, $\theta^{N^*}$ denotes the discrete function defined on the fine grid, and $\mathcal{I}(\theta^{N})$ denotes the discrete function obtained interpolating linearly the discrete function $\theta^{N}$ into the fine mesh. The global space-time relative difference is computed as:
\begin{equation}
   \text{D}^{N} \theta = \frac{\sum_{i,k\in \mathcal{S},\mathcal{T}}\left| \mathcal{I}(\theta^{N})(x_i,t_k)-\theta^{N^*}(x_i,t_k)\right|}{\sum_{i,k\in \mathcal{S},\mathcal{T}}\left| \theta^{N^*}(x_i,t_k)\right|}~,\label{eq:global-rel-diff}
\end{equation}
where $\mathcal{T}$ refers to the temporal points indices.

\begin{figure}[h]
	\centering
	\includegraphics[width=0.95\textwidth]{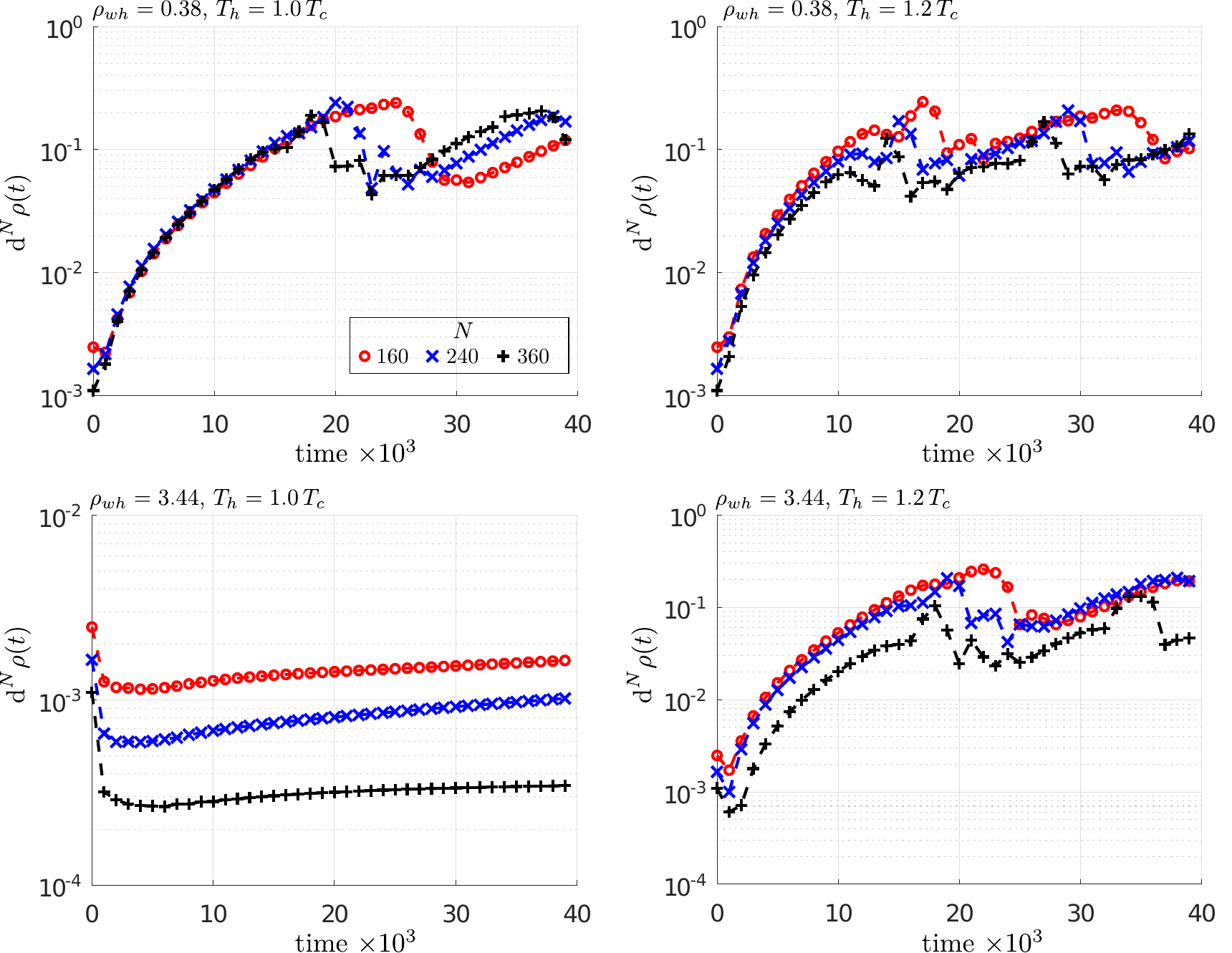}
	\caption{Relative differences in space per timestep, computed according to Eq. \eqref{eq:rel-diff-timestep} for the density field along three pairs of refinements $(N,N^*)\in\{(160,240),(240,360),(360,540)\}$. $\rho_{wh}$ and $T_h$ denote the $\psi$-based contact angle parameter and the temperature set at the heater lattices respectively.}\label{fig:pool-relative-diff-pertimesttep_RHO}
\end{figure}

\begin{figure}[h]
	\centering
	\includegraphics[width=0.95\textwidth]{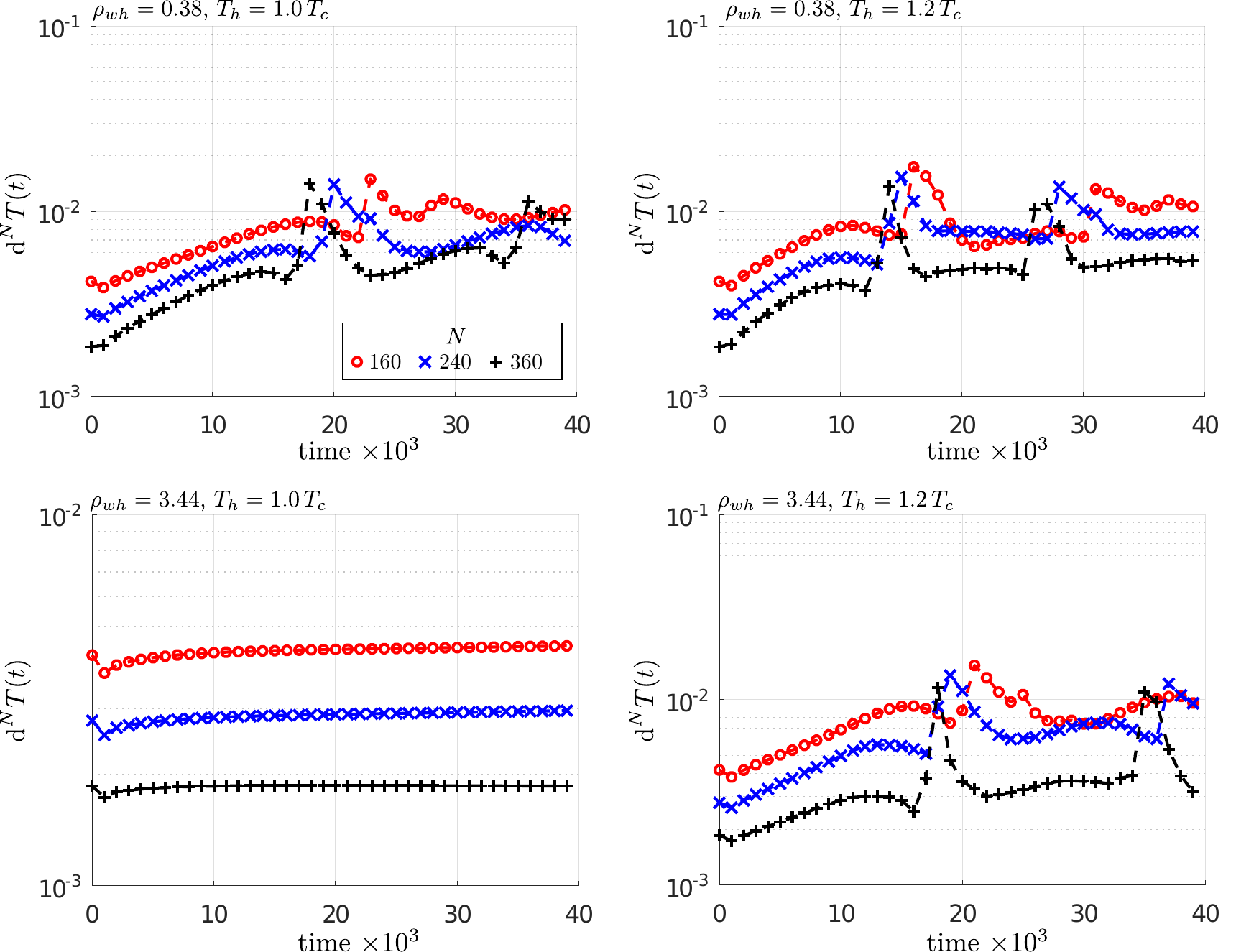}
	\caption{Relative differences in space per timestep, computed according to Eq. \eqref{eq:rel-diff-timestep} for the temperature field along three pairs of refinements $(N,N^*)\in\{(160,240),(240,360),(360,540)\}$. $\rho_{wh}$ and $T_h$ denote the $\psi$-based contact angle parameter and the temperature set at the heater lattices respectively.}\label{fig:pool-relative-diff-pertimesttep_T}
\end{figure}

The simulations run a number of timesteps such that the first bubble have departed and the second bubble have been formed for three of the four configurations tested here (more precisely, running along 40000 timesteps for the coarsest mesh of $160\times 160$ lattices). The case $\rho_{wh}=3.44$ and $T_h=1.0\,T_c$ corresponds to a configuration where nucleation is not observed. The relative spatial differences for density and temperature, computed according to Eq. \eqref{eq:rel-diff-timestep}, are shown in Fig. \ref{fig:pool-relative-diff-pertimesttep_RHO} and \ref{fig:pool-relative-diff-pertimesttep_T}, respectively. As expected, the spatial relative difference grows in time, and the refinement of the meshes (augmenting the number of lattices and recomputing the physical parameters as discussed in Section \ref{sec:physics-meshes}) has the effect of decreasing the values of these differences in time. This happens more clearly in the case without nucleation.

The overall order of convergence is obtained by computing the global error given by Eq. \eqref{eq:global-rel-diff}. The resulting global errors for each configuration and the respective orders of convergence are shown in Fig. \ref{fig:pool-relative-diff-global}, left column. It is worth noticing that the convergence order depends on the parameters of the simulation and the considered field. For instance, fixing $\rho_{wh}=3.44$, one observes a convergence ratio of order 1 for the temperature field, independently of the heater temperature, while for the same value of $\rho_{wh}$ the convergence ratio for the density field is 1.2 for $T_{h}=1.2\,T_c$ and 1.8 for the lower temperature $T_{h}=1.0\,T_c$. The case $\rho_{wh}=0.38$ and $T_{h}=1.0\,T_c$ is more complex. As it can be observed, the relative differences between meshes for the density field presented almost no change for that case, diminishing with a ratio of $\approx 0.06$. However, below, we show that this low convergence order is due to the delay of the simulations between meshes detected in Section \ref{sec:numerics-boil-qualitative-comparisons}.

\subsubsection*{Time-shifting the simulations}
Instead of evaluating the relative difference employing Eq. \eqref{eq:global-rel-diff}, given two meshes and two shift parameters $\delta$ and $\delta^*$, the shifted relative difference is computed as:
\begin{equation}
   \text{D}^{N;\delta,\delta^*} \theta = \frac{\sum_{i,k\in \mathcal{S},\mathcal{T}}\left| \mathcal{I}(\theta^{N})\left(x_i,t_k+\frac{\delta}{\tau} t_k\right)-\theta^{N^*}\left(x_i,t_k+\frac{\delta^*}{\tau} t_k\right)\right|}{\sum_{i,k\in \mathcal{S},\mathcal{T}}\left| \theta^{N^*}\left(x_i,t_k+\frac{\delta^*}{\tau} t_k\right)\right|}~,\label{eq:global-rel-diff-shifted}
\end{equation}
where $\delta$ ($\delta^*$) is the shift necessary to \emph{synchronize}, at a certain timestep denoted by $\tau$, the mesh $N$ ($N^*$) with the finest mesh, which corresponds to $540\times 540$ lattices. For a given mesh with $N\times N$ lattices, these $\delta$'s are computed by solving a one-dimensional minimization problem:
\begin{equation}
\delta =\underset{-m \leq \delta \leq m}{\argmin} ~
\sum_{i\in \mathcal{S}}\left| \mathcal{I}(\theta^{N})(x_i,\tau+\delta)-\theta^{\text{finest}}(x_i,\tau)\right|~,\label{eq:rel-diff-minimize}
\end{equation}
where $m$ defines the searching interval size. For the four configurations tested the reference time was set to $\tau=5120$ and the size of the search interval size was set to $m=4800$, obtaining always $|\delta|<m$, which warranties this $\delta$ is defined in terms of a local minimum. The values of $\delta$ obtained for each case and mesh (relative to the finest mesh) are given in Table \ref{tab:pool-shifts}. Notice that, as expected, the values of $\delta$ diminish as the mesh is refined, which is an indicator of local convergence in time. The case $\rho_{wh}=3.44$ and $T_{h}=1.0\, T_c$ presented no nucleation and for $t=\tau$ the system already reached a stationary regime, which had as a consequence that $\delta$ resulted to be null for the coarser meshes. 

\begin{table}[b]
\centering
\begin{tabular}{c|cccc}
\toprule
 &   \multicolumn{2} {c} {$\rho_{wh}=0.38$} &  \multicolumn{2} {c} {$\rho_{wh}=3.44$} \\
 &  $T_{h}/T_c=1.0$ & $T_{h}/T_c=1.2$ & $T_{h}/T_c=1.0$  & $T_{h}/T_c=1.2$ \\
\midrule
$N=160$ & 6720 & 3840 & 0 & 4480\\

$N=240$ & 2880 & 1600 & 0 & 1600\\

$N=360$ & 960 & 640 & 0  & 320\\
\bottomrule
\end{tabular}\caption{Values of the time-shifts $\delta$ computed by Eq. \eqref{eq:rel-diff-minimize} to synchronize the simulations of each coarse mesh with the finest one at the time $\tau=5120$.}
\label{tab:pool-shifts}
\end{table}

\begin{figure}[h]
	\centering
	\includegraphics[width=0.9\textwidth]{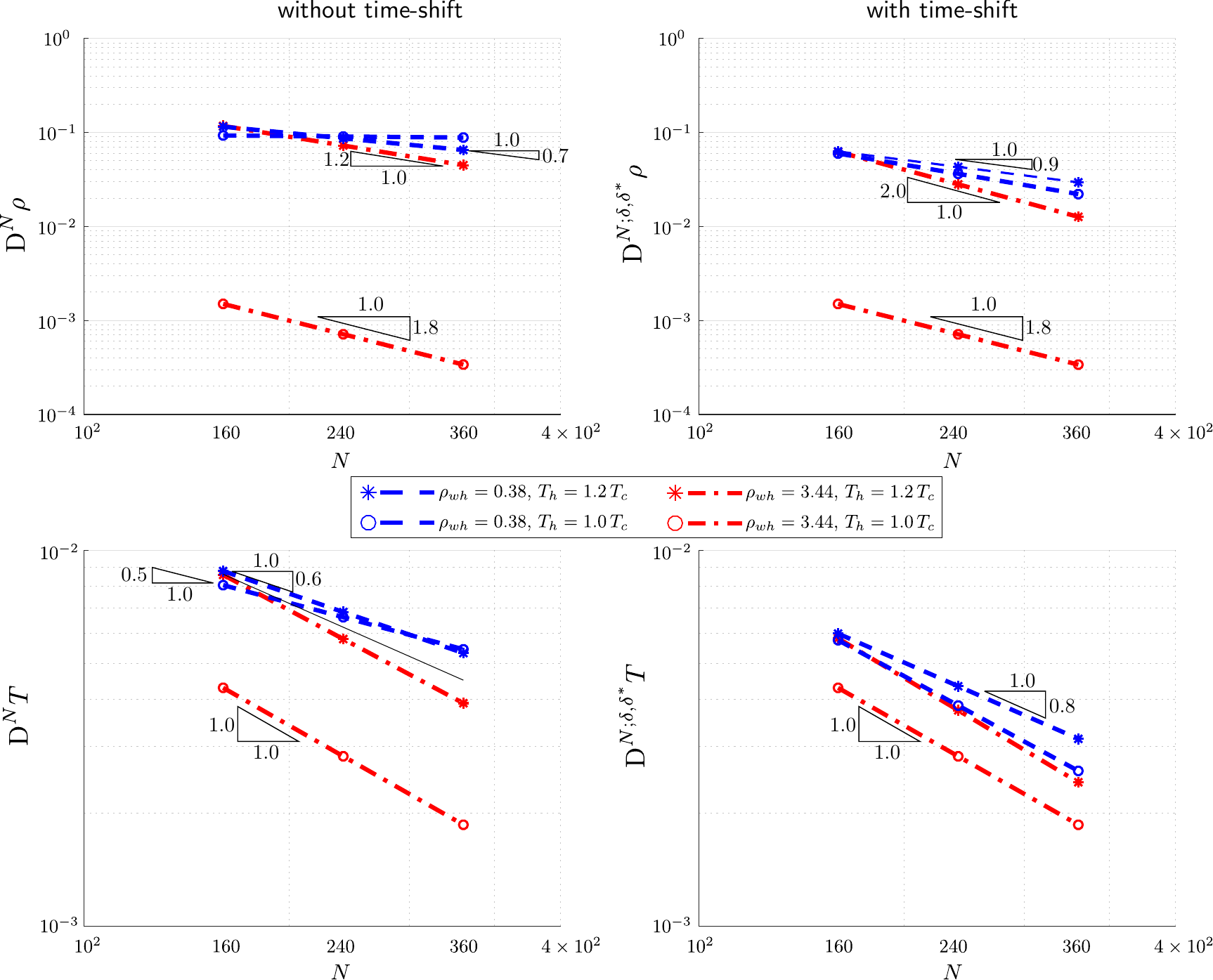}
	\caption{Global relative differences for the pool boiling according to Eq. \eqref{eq:global-rel-diff} (left side) and to Eqs. \eqref{eq:global-rel-diff-shifted}-\eqref{eq:rel-diff-minimize} (right side).}\label{fig:pool-relative-diff-global}
\end{figure}

The resulting global relative differences with these time-shifts are shown at the right side of Fig. \ref{fig:pool-relative-diff-global}. First, it is worth noticing that the trend of the global differences among meshes is to diminish when the shifting is applied, where the only exception is the case where no nucleation takes place (a case for which $\delta$ is 0). Regarding the temperature field, the convergence order was around the unit for all the cases tested. On the other hand, all the cases where nucleation takes place exhibit a convergence order around the unit for the density field, while for the case where nucleation is not observed the convergence order is $\approx 2$.
These results are evidence of time and space convergence for the pool boiling problem. As the relative differences employed here summarize the variations along with both time and space in norm $L^1$, these differences going to zero also represent evidence of convergence for a series of quantities depending on $\rho$, like bubbles' geometry at a certain time, departure time and contact angle. 

\begin{figure}[hb!]
	\centering
	\includegraphics[width=\textwidth]{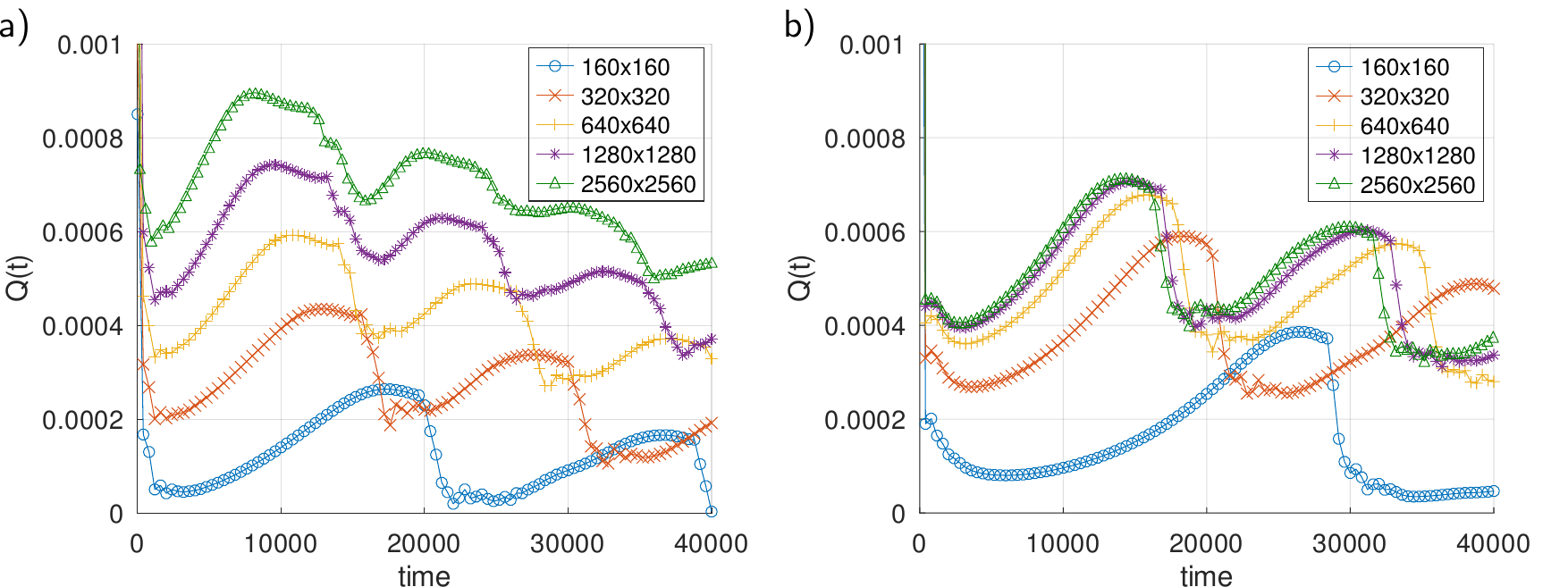}
	\caption{Heat fluxes calculated for four different meshes considering $T_h=1.2$. a) $\rho_{wh}=0.38$ and b) $\rho_{wh}=3.44$.}\label{fig:pool-heat-flux}
\end{figure}

A quantity of engineering interest that depends both on $\rho$ and $T$ is the average heat flux entering the computational domain, defined as
\begin{equation}
    Q(t) = \frac{1}{L}\int_{0}^L \left(\lambda(\rho(x,y,t)) \parder{T}{y}(x,y,t)\right)_{y=0}\,dx ~.
    \label{eq:pool-heat-flux}
\end{equation}
This quantity is shown in Fig. \ref{fig:pool-heat-flux} for $T_h=1.2$ and two values of the contact angle parameter at the heater lattices: a) $\rho_{wh}=0.38$, and b) $\rho_{wh}=3.44$. For these cases we have changed the mesh refinement ratio to 2.0 and we have added a fifth mesh of $2560\times 2560$ lattices. For $\rho_{wh}=3.44$, taking the relative differences of the curves in time, resulted in a convergence order of around 1.0. While for $\rho_{wh}=0.38$ the convergence order was lower, around 0.5. This behavior indicates that this ratio is very sensitive to the contact angle configurations. Also, from these curves, one can observe the delay taking place between the simulations and how they tend to synchronize as the number of computational lattices is augmented. For the refinement ratio of 1.5 the same behavior was observed for both cases, and we choose to show the more refined results.

With the mesh refinement we can also observe that the number of nucleated bubbles in time increases for finer meshes, resulting in three bubble departures for the finest mesh when the heater contact angle parameter is $\rho_{wh}=0.38$ (see Fig. \ref{fig:pool-heat-flux} a). In this case the number of nucleated bubbles raised from two to three, while for $\rho_{wh}=0.38$ the number of nucleated bubbles increase from one to two (see Fig. \ref{fig:pool-heat-flux} b). This fact, which is related to the previously commented time delay for different mesh sizes, shows the importance of reaching numerically converged solutions in order to correctly describe the dynamics of the boiling process. Moreover, one observes that the increase of the bubbles departure frequency is associated with a higher overall heat flux, which is a well known physical behavior.


\FloatBarrier

\section{Conclusions}
\label{sec:conclusions}

The following conclusions can be drawn from the present work:
\begin{itemize}
    \item An original mesh refinement procedure has been proposed to obtain mesh independent simulation results with pseudopotential LBM of phase-change heat transfer problems, specially pool boiling problems, in the viscous regime. The procedure was based on conserving the hydrodynamic and thermodynamic similarities between different meshes by suitably adjusting the parameters $a$, $\sigma$, $\kappa$ and $g$.
    \item It has been shown that the attainment of numerically converged results of pool boiling depends on both the choice of the contact angle model and its tuning parameters ($G_w$ for the modified-$\psi$-based model  and $\rho_w$ for the $\psi$-based model). The $\psi$-based model was the method that produced consistent results without requiring the adjustment of its defining parameter.
    \item A delay of pool boiling converged simulation results for different meshes has been observed. This could affect quantities of engineering interest like the bubbles' frequency departure, spatial average heat flux and boiling curves.  
    \item The pool boiling and the Stefan problems results suggest that the convergence order is affected by the presence of nucleation and the compressibility of the fluid. This shows that the incorporation of more complex physics into the simulations affects the convergence ratio and thus would require more computational effort to well describe the physical phenomena occurring in the simulated problems.
    \item The proposed procedure and consequently the simulation of numerically converged results required an extensive use of computational resources, indicating the necessity of using LBM parallel numerical codes for this purpose.
    \item The results obtained suggest the importance of carrying out mesh refinement in order to correctly assess pool boiling problems with the LBM. 
\end{itemize}

\section*{Acknowledgments}

The authors acknowledge the National Laboratory for Scientific Computing (LNCC/MCTI, Brazil) for providing HPC resources of the SDumont supercomputer, which have contributed to the research results reported within this paper.

Part of this research was carried out using the computational resources of the Center for Mathematical Sciences Applied to Industry (CeMEAI) funded by FAPESP (grant 2013/07375-0). 

The authors acknowledge the support received from the São Paulo Foundation for Research Support (FAPESP), grants 2016/09509-1, 2020/12919-2, and the National Council for Scientific and Technological Development (CNPq), grants 431782/2018-0, 140634/2019-3, 305941/2020-8.

\section*{Appendix A}
\label{sec:appendix}
\setcounter{table}{0}
\renewcommand{\thetable}{A\arabic{table}}

In this appendix some details on the computational machines and the execution times are provided. 

An in-house code was developed in the programming language C along with the parallelization paradigm of the Message Passing Interface (MPI).

The simulations for the heat flux curves provided in Section \ref{sec:numerics-pool-boiling-results} were executed in the \emph{Euler Supercomputer}\footnote{\url{https://euler.cemeai.icmc.usp.br}},
that is located at the Institute of Mathematical and Computer Sciences (ICMC) of the University of
S\~ao Paulo at S\~ao Carlos. All the other simulations were executed in the cluster Santos Dumont, located at the National Laboratory for Scientific Computing (LNCC) \footnote{\url{https://sdumont.lncc.br}}. The hardware specifications of both clusters are listed in Table \ref{tab:computational_nodes}.
\begin{table}[h!]
	\begin{center}
		\begin{tabular}{lll}
			\toprule
			 & \emph{Cluster Sdumont}  & \emph{Cluster Euler}     \\ 
			\midrule
			Processors & 2x Intel Xeon Cascade Lake Gold 6252 & 2x Intel Xeon \verb|E5-2680v2|    \\
			Cores per processor  & 24 (48 per node)  & 10 (20 per nodes)\\
			RAM per node &  384 Gb & 128 Gb \\
                        Communications & InfiniBand & InfiniBand\\
                        \bottomrule
		\end{tabular}
		\caption{Specifications of the computational nodes were the simulations were executed.}\label{tab:computational_nodes}
	\end{center}
\end{table}
\begin{table}[h!]
	\begin{center}
		\begin{tabular}{crcc}
			\toprule
		Mesh & Execution time  & \# Cores used & Mlups   \\ 
			\midrule
			$160\times160$ & 0 m 37 s & 40 & 30 \\
			$320\times 320$  & 4 m 21 s & 80 & 69 \\
			$640\times 640$ & 45 m 35 s & 80 & 105 \\
            $1280\times 1280$ & 391 m 36 s & 160 & 196 \\
            $2560\times 2560$ & 4135 m 30 s & 256 & 297 \\
                        \bottomrule
		\end{tabular}
		\caption{Some efficiency data of the in-house code during the heat flux simulations.}\label{tab:computational_costs}
	\end{center}
\end{table}

For completeness, the computational times needed to execute each of the 8 simulations summarized in Fig. \ref{fig:pool-heat-flux} are given in Table \ref{tab:computational_costs}. These computational runs were performed in the Supercomputer Euler where the Intel Suite Compilers was employed. Observe that, as it is typical in High Performace Computations, the more refined the mesh is the more efficient the code execution. In fact, for the smaller mesh our delivers around 0.75 million lattice updates per second (Mlups) per core, while for the finer mesh the efficiency is of about 1.23 Mlups per core. 

\FloatBarrier


\begin{thebibliography}{78}
\expandafter\ifx\csname natexlab\endcsname\relax\def\natexlab#1{#1}\fi
\providecommand{\url}[1]{\texttt{#1}}
\providecommand{\href}[2]{#2}
\providecommand{\path}[1]{#1}
\providecommand{\DOIprefix}{doi:}
\providecommand{\ArXivprefix}{arXiv:}
\providecommand{\URLprefix}{URL: }
\providecommand{\Pubmedprefix}{pmid:}
\providecommand{\doi}[1]{\href{http://dx.doi.org/#1}{\path{#1}}}
\providecommand{\Pubmed}[1]{\href{pmid:#1}{\path{#1}}}
\providecommand{\bibinfo}[2]{#2}
\ifx\xfnm\relax \def\xfnm[#1]{\unskip,\space#1}\fi
\bibitem[{Alexiades et~al.(1993)Alexiades, Solomon and
  Lunardini}]{Alexiades1993}
\bibinfo{author}{Alexiades, V.}, \bibinfo{author}{Solomon, A.D.},
  \bibinfo{author}{Lunardini, V.J.}, \bibinfo{year}{1993}.
\newblock \bibinfo{title}{{Mathematical Modeling of Melting and Freezing
  Processes}}.
\newblock \bibinfo{publisher}{Taylor \& Francis Group}.
\bibitem[{Benzi et~al.(2006)Benzi, Biferale, Sbragaglia, Succi and
  Toschi}]{Benzi2006}
\bibinfo{author}{Benzi, R.}, \bibinfo{author}{Biferale, L.},
  \bibinfo{author}{Sbragaglia, M.}, \bibinfo{author}{Succi, S.},
  \bibinfo{author}{Toschi, F.}, \bibinfo{year}{2006}.
\newblock \bibinfo{title}{{Mesoscopic modeling of a two-phase flow in the
  presence of boundaries: The contact angle}}.
\newblock \bibinfo{journal}{Physical Review E} \bibinfo{volume}{74},
  \bibinfo{pages}{021509}.
\bibitem[{Bhatnagar et~al.(1954)Bhatnagar, Gross and
  Krook}]{Bhatnagar_1954_ModelCollisionProcesses}
\bibinfo{author}{Bhatnagar, P.L.}, \bibinfo{author}{Gross, E.P.},
  \bibinfo{author}{Krook, M.}, \bibinfo{year}{1954}.
\newblock \bibinfo{title}{A model for collision processes in gases. i. small
  amplitude processes in charged and neutral one-component systems}.
\newblock \bibinfo{journal}{Physical Review} \bibinfo{volume}{94},
  \bibinfo{pages}{511--525}.
\bibitem[{Chen et~al.(2004)Chen, Orszag, Staroselsky and
  Succi}]{Chen_2004_ExpandedanalogyBoltzmann}
\bibinfo{author}{Chen, H.}, \bibinfo{author}{Orszag, S.A.},
  \bibinfo{author}{Staroselsky, I.}, \bibinfo{author}{Succi, S.},
  \bibinfo{year}{2004}.
\newblock \bibinfo{title}{Expanded analogy between {B}oltzmann kinetic theory
  of fluids and turbulence}.
\newblock \bibinfo{journal}{Journal of Fluid Mechanics} \bibinfo{volume}{519},
  \bibinfo{pages}{301--314}.
\bibitem[{Chen and Doolen(1998)}]{Chen_1998_LatticeBoltzmannmethod}
\bibinfo{author}{Chen, S.}, \bibinfo{author}{Doolen, G.D.},
  \bibinfo{year}{1998}.
\newblock \bibinfo{title}{Lattice {B}oltzmann method for fluid flows}.
\newblock \bibinfo{journal}{Annual review of fluid mechanics}
  \bibinfo{volume}{30}, \bibinfo{pages}{329--364}.
\bibitem[{Clerk-Maxwell(1875)}]{clerk1875dynamical}
\bibinfo{author}{Clerk-Maxwell, J.}, \bibinfo{year}{1875}.
\newblock \bibinfo{title}{On the dynamical evidence of the molecular
  constitution of bodies}.
\newblock \bibinfo{journal}{Nature} \bibinfo{volume}{11},
  \bibinfo{pages}{357--359}.
\bibitem[{Czelusniak et~al.(2020)Czelusniak, Mapelli, Guzella,
  Cabezas-G{\'{o}}mez and Wagner}]{Czelusniak2020}
\bibinfo{author}{Czelusniak, L.E.}, \bibinfo{author}{Mapelli, V.P.},
  \bibinfo{author}{Guzella, M.S.}, \bibinfo{author}{Cabezas-G{\'{o}}mez, L.},
  \bibinfo{author}{Wagner, A.J.}, \bibinfo{year}{2020}.
\newblock \bibinfo{title}{{Force approach for the pseudopotential lattice
  Boltzmann method}}.
\newblock \bibinfo{journal}{Physical Review E} \bibinfo{volume}{102},
  \bibinfo{pages}{033307}.
\bibitem[{{De Bortoli} et~al.(2015){De Bortoli}, Andreis and
  Pereira}]{Debortoli2015}
\bibinfo{author}{{De Bortoli}, A.L.}, \bibinfo{author}{Andreis, G.S.},
  \bibinfo{author}{Pereira, F.N.}, \bibinfo{year}{2015}.
\newblock \bibinfo{title}{Modeling and Simulation of Reactive Flows}.
\newblock \bibinfo{publisher}{Elsevier}.
\bibitem[{d'Humi{\`e}res(1992)}]{d1992generalized}
\bibinfo{author}{d'Humi{\`e}res, D.}, \bibinfo{year}{1992}.
\newblock \bibinfo{title}{Generalized lattice-{B}oltzmann equations}.
\newblock \bibinfo{journal}{Rarefied gas dynamics} .
\bibitem[{Dong et~al.(2018)Dong, Gong and Cheng}]{Dong2018}
\bibinfo{author}{Dong, L.}, \bibinfo{author}{Gong, S.}, \bibinfo{author}{Cheng,
  P.}, \bibinfo{year}{2018}.
\newblock \bibinfo{title}{Direct numerical simulations of film boiling heat
  transfer by a phase-change lattice boltzmann method}.
\newblock \bibinfo{journal}{International Communications in Heat and Mass
  Transfer} \bibinfo{volume}{91}, \bibinfo{pages}{109--116}.
\bibitem[{Fari{\~{n}}as et~al.(2019)Fari{\~{n}}as, S{\'{a}}nchez, Guzella,
  Amado, S{\'{a}}iz and {Cabezas-G{\'{o}}mez}}]{Farinas2019}
\bibinfo{author}{Fari{\~{n}}as, P.}, \bibinfo{author}{S{\'{a}}nchez, M.L.},
  \bibinfo{author}{Guzella, M.}, \bibinfo{author}{Amado, J.M.},
  \bibinfo{author}{S{\'{a}}iz, J.M.}, \bibinfo{author}{{Cabezas-G{\'{o}}mez},
  L.}, \bibinfo{year}{2019}.
\newblock \bibinfo{title}{{Experimental investigation of the CHF of HFE-7100
  under pool boiling conditions on differently roughened surfaces}}.
\newblock \bibinfo{journal}{International Journal of Heat and Mass Transfer}
  \bibinfo{volume}{139}, \bibinfo{pages}{269--279}.
\bibitem[{Fei et~al.(2022)Fei, Qin, Wang, Luo, Derome and Carmeliet}]{Fei2022}
\bibinfo{author}{Fei, L.}, \bibinfo{author}{Qin, F.}, \bibinfo{author}{Wang,
  G.}, \bibinfo{author}{Luo, K.H.}, \bibinfo{author}{Derome, D.},
  \bibinfo{author}{Carmeliet, J.}, \bibinfo{year}{2022}.
\newblock \bibinfo{title}{{Droplet evaporation in finite-size systems:
  Theoretical analysis and mesoscopic modeling}}.
\newblock \bibinfo{journal}{Physical Review E} \bibinfo{volume}{105},
  \bibinfo{pages}{025101}.
\bibitem[{Fei et~al.(2020)Fei, Yang, Chen, Mo and Luo}]{Fei2020}
\bibinfo{author}{Fei, L.}, \bibinfo{author}{Yang, J.}, \bibinfo{author}{Chen,
  Y.}, \bibinfo{author}{Mo, H.}, \bibinfo{author}{Luo, K.H.},
  \bibinfo{year}{2020}.
\newblock \bibinfo{title}{{Mesoscopic simulation of three-dimensional pool
  boiling based on a phase-change cascaded lattice Boltzmann method}}.
\newblock \bibinfo{journal}{Physics of Fluids} \bibinfo{volume}{32},
  \bibinfo{pages}{103312}.
\bibitem[{Feng et~al.(2021a)Feng, Chang, Hu, Li and Zhao}]{Feng2021}
\bibinfo{author}{Feng, Y.}, \bibinfo{author}{Chang, F.}, \bibinfo{author}{Hu,
  Z.}, \bibinfo{author}{Li, H.}, \bibinfo{author}{Zhao, J.},
  \bibinfo{year}{2021}a.
\newblock \bibinfo{title}{Investigation of pool boiling heat transfer on
  hydrophilic-hydrophobic mixed surface with micro-pillars using lbm}.
\newblock \bibinfo{journal}{International Journal of Thermal Sciences}
  \bibinfo{volume}{163}, \bibinfo{pages}{106814}.
\bibitem[{Feng et~al.(2019a)Feng, Li, Guo, Lei and Zhao}]{Feng2019b}
\bibinfo{author}{Feng, Y.}, \bibinfo{author}{Li, H.}, \bibinfo{author}{Guo,
  K.}, \bibinfo{author}{Lei, X.}, \bibinfo{author}{Zhao, J.},
  \bibinfo{year}{2019}a.
\newblock \bibinfo{title}{Numerical investigation on bubble dynamics during
  pool nucleate boiling in presence of a non-uniform electric field by lbm}.
\newblock \bibinfo{journal}{Applied Thermal Engineering} \bibinfo{volume}{155},
  \bibinfo{pages}{637--649}.
\bibitem[{Feng et~al.(2019b)Feng, Li, Guo, Lei and Zhao}]{Feng2019a}
\bibinfo{author}{Feng, Y.}, \bibinfo{author}{Li, H.}, \bibinfo{author}{Guo,
  K.}, \bibinfo{author}{Lei, X.}, \bibinfo{author}{Zhao, J.},
  \bibinfo{year}{2019}b.
\newblock \bibinfo{title}{Numerical study on saturated pool boiling heat
  transfer in presence of a uniform electric field using lattice boltzmann
  method}.
\newblock \bibinfo{journal}{International Journal of Heat and Mass Transfer}
  \bibinfo{volume}{135}, \bibinfo{pages}{885--896}.
\bibitem[{Feng et~al.(2021b)Feng, Li, Guo, Lei and Zhao}]{Feng2021b}
\bibinfo{author}{Feng, Y.}, \bibinfo{author}{Li, H.}, \bibinfo{author}{Guo,
  K.}, \bibinfo{author}{Lei, X.}, \bibinfo{author}{Zhao, J.},
  \bibinfo{year}{2021}b.
\newblock \bibinfo{title}{{Lattice Boltzmann Study on Influence of
  Gravitational Acceleration on Pool Nucleate Boiling Heat Transfer}}.
\newblock \bibinfo{journal}{Microgravity Science and Technology}
  \bibinfo{volume}{33}, \bibinfo{pages}{21}.
\bibitem[{Feng et~al.(2018)Feng, Li, Guo, Zhao and Wang}]{Feng2018}
\bibinfo{author}{Feng, Y.}, \bibinfo{author}{Li, H.}, \bibinfo{author}{Guo,
  K.}, \bibinfo{author}{Zhao, J.}, \bibinfo{author}{Wang, T.},
  \bibinfo{year}{2018}.
\newblock \bibinfo{title}{{Numerical Study of Single Bubble Growth on and
  Departure from a Horizontal Superheated Wall by Three-dimensional Lattice
  Boltzmann Method}}.
\newblock \bibinfo{journal}{Microgravity Science and Technology}
  \bibinfo{volume}{30}, \bibinfo{pages}{761--773}.
\bibitem[{Gong and Cheng(2012)}]{Gong_2012_Numericalinvestigationdroplet}
\bibinfo{author}{Gong, S.}, \bibinfo{author}{Cheng, P.}, \bibinfo{year}{2012}.
\newblock \bibinfo{title}{Numerical investigation of droplet motion and
  coalescence by an improved lattice {B}oltzmann model for phase transitions
  and multiphase flows}.
\newblock \bibinfo{journal}{Computers \& Fluids} \bibinfo{volume}{53},
  \bibinfo{pages}{93--104}.
\bibitem[{Gong and Cheng(2013)}]{Gong2013}
\bibinfo{author}{Gong, S.}, \bibinfo{author}{Cheng, P.}, \bibinfo{year}{2013}.
\newblock \bibinfo{title}{{Lattice Boltzmann simulation of periodic bubble
  nucleation, growth and departure from a heated surface in pool boiling}}.
\newblock \bibinfo{journal}{International Journal of Heat and Mass Transfer}
  \bibinfo{volume}{64}, \bibinfo{pages}{122--132}.
\bibitem[{Gong and Cheng(2015a)}]{Gong_2015_LatticeBoltzmannsimulations}
\bibinfo{author}{Gong, S.}, \bibinfo{author}{Cheng, P.}, \bibinfo{year}{2015}a.
\newblock \bibinfo{title}{Lattice {B}oltzmann simulations for surface
  wettability effects in saturated pool boiling heat transfer}.
\newblock \bibinfo{journal}{International Journal of Heat and Mass Transfer}
  \bibinfo{volume}{85}, \bibinfo{pages}{635--646}.
\bibitem[{Gong and Cheng(2015b)}]{Gong_2015_Numericalsimulationpool}
\bibinfo{author}{Gong, S.}, \bibinfo{author}{Cheng, P.}, \bibinfo{year}{2015}b.
\newblock \bibinfo{title}{Numerical simulation of pool boiling heat transfer on
  smooth surfaces with mixed wettability by lattice {B}oltzmann method}.
\newblock \bibinfo{journal}{International Journal of Heat and Mass Transfer}
  \bibinfo{volume}{80}, \bibinfo{pages}{206--216}.
\bibitem[{Gong and Cheng(2017)}]{Gong_2017_Directnumericalsimulations}
\bibinfo{author}{Gong, S.}, \bibinfo{author}{Cheng, P.}, \bibinfo{year}{2017}.
\newblock \bibinfo{title}{Direct numerical simulations of pool boiling curves
  including heater's thermal responses and the effect of vapor phase's thermal
  conductivity}.
\newblock \bibinfo{journal}{International Communications in Heat and Mass
  Transfer} \bibinfo{volume}{87}, \bibinfo{pages}{61--71}.
\bibitem[{Gong et~al.(2018)Gong, Yan, Chen and
  Wright}]{Gong_2018_modifiedphasechange}
\bibinfo{author}{Gong, W.}, \bibinfo{author}{Yan, Y.Y.}, \bibinfo{author}{Chen,
  S.}, \bibinfo{author}{Wright, E.}, \bibinfo{year}{2018}.
\newblock \bibinfo{title}{A modified phase change pseudopotential lattice
  {B}oltzmann model}.
\newblock \bibinfo{journal}{International Journal of Heat and Mass Transfer}
  \bibinfo{volume}{125}, \bibinfo{pages}{323--329}.
\bibitem[{Guan et~al.(2011)Guan, Klausner and Mei}]{Guan2011}
\bibinfo{author}{Guan, C.K.}, \bibinfo{author}{Klausner, J.F.},
  \bibinfo{author}{Mei, R.}, \bibinfo{year}{2011}.
\newblock \bibinfo{title}{A new mechanistic model for pool boiling chf on
  horizontal surfaces}.
\newblock \bibinfo{journal}{International Journal of Heat and Mass Transfer}
  \bibinfo{volume}{54}, \bibinfo{pages}{3960--3969}.
\bibitem[{Gunstensen et~al.(1991)Gunstensen, Rothman, Zaleski and
  Zanetti}]{Gunstensen_1991_LatticeBoltzmannmodel}
\bibinfo{author}{Gunstensen, A.K.}, \bibinfo{author}{Rothman, D.H.},
  \bibinfo{author}{Zaleski, S.}, \bibinfo{author}{Zanetti, G.},
  \bibinfo{year}{1991}.
\newblock \bibinfo{title}{Lattice {B}oltzmann model of immiscible fluids}.
\newblock \bibinfo{journal}{Physical Review A} \bibinfo{volume}{43},
  \bibinfo{pages}{4320--4327}.
\bibitem[{Guo and Zhao(2002)}]{Guo_2002_LatticeBoltzmannmodel}
\bibinfo{author}{Guo, Z.}, \bibinfo{author}{Zhao, T.S.}, \bibinfo{year}{2002}.
\newblock \bibinfo{title}{Lattice {B}oltzmann model for incompressible flows
  through porous media}.
\newblock \bibinfo{journal}{Physical Review E} \bibinfo{volume}{66},
  \bibinfo{pages}{036304}.
\bibitem[{Guo et~al.(2002)Guo, Zheng and Shi}]{Guo_2002_Discretelatticeeffects}
\bibinfo{author}{Guo, Z.}, \bibinfo{author}{Zheng, C.}, \bibinfo{author}{Shi,
  B.}, \bibinfo{year}{2002}.
\newblock \bibinfo{title}{Discrete lattice effects on the forcing term in the
  lattice {B}oltzmann method}.
\newblock \bibinfo{journal}{Physical Review E} \bibinfo{volume}{65},
  \bibinfo{pages}{046308}.
\bibitem[{Guzella et~al.(2020)Guzella, Czelusniak, Mapelli, Fari{\~n}as,
  Ribatski and Cabezas-G{\'o}mez}]{guzella2020simulation}
\bibinfo{author}{Guzella, M.}, \bibinfo{author}{Czelusniak, L.E.},
  \bibinfo{author}{Mapelli, V.}, \bibinfo{author}{Fari{\~n}as, P.},
  \bibinfo{author}{Ribatski, G.}, \bibinfo{author}{Cabezas-G{\'o}mez, L.},
  \bibinfo{year}{2020}.
\newblock \bibinfo{title}{Simulation of boiling heat transfer at different
  reduced temperatures with an improved pseudopotential lattice boltzmann
  method}.
\newblock \bibinfo{journal}{Symmetry} \bibinfo{volume}{12},
  \bibinfo{pages}{1358}.
\bibitem[{Haramura and Katto(1983)}]{Haramura1983}
\bibinfo{author}{Haramura, Y.}, \bibinfo{author}{Katto, Y.},
  \bibinfo{year}{1983}.
\newblock \bibinfo{title}{A new hydrodynamic model of critical heat flux,
  applicable widely to both pool and forced convection boiling on submerged
  bodies in saturated liquids}.
\newblock \bibinfo{journal}{International Journal of Heat and Mass Transfer}
  \bibinfo{volume}{26}, \bibinfo{pages}{389--399}.
\bibitem[{Higuera and
  Jim{\'e}nez(1989)}]{Higuera_1989_Boltzmannapproachlattice}
\bibinfo{author}{Higuera, F.J.}, \bibinfo{author}{Jim{\'e}nez, J.},
  \bibinfo{year}{1989}.
\newblock \bibinfo{title}{{B}oltzmann approach to lattice gas simulations}.
\newblock \bibinfo{journal}{Europhysics Letters} \bibinfo{volume}{9},
  \bibinfo{pages}{663}.
\bibitem[{Huang et~al.(2009)Huang, Li, Liu and Lu}]{Huang2009}
\bibinfo{author}{Huang, H.}, \bibinfo{author}{Li, Z.}, \bibinfo{author}{Liu,
  S.}, \bibinfo{author}{Lu, X.y.}, \bibinfo{year}{2009}.
\newblock \bibinfo{title}{{Shan-and-Chen-type multiphase lattice Boltzmann
  study of viscous coupling effects for two-phase flow in porous media}}.
\newblock \bibinfo{journal}{International Journal for Numerical Methods in
  Fluids} \bibinfo{volume}{61}, \bibinfo{pages}{341--354}.
\bibitem[{Kandlikar(2001)}]{Kandlikar2001}
\bibinfo{author}{Kandlikar, S.G.}, \bibinfo{year}{2001}.
\newblock \bibinfo{title}{{A Theoretical Model to Predict Pool Boiling CHF
  Incorporating Effects of Contact Angle and Orientation }}.
\newblock \bibinfo{journal}{Journal of Heat Transfer} \bibinfo{volume}{123},
  \bibinfo{pages}{1071--1079}.
\bibitem[{Kharmiani et~al.(2019)Kharmiani, Niazmand and
  Passandideh-Fard}]{Kharmiani_2019_AlternativeHighDensity}
\bibinfo{author}{Kharmiani, S.F.}, \bibinfo{author}{Niazmand, H.},
  \bibinfo{author}{Passandideh-Fard, M.}, \bibinfo{year}{2019}.
\newblock \bibinfo{title}{An alternative high-density ratio pseudo-potential
  lattice {B}oltzmann model with surface tension adjustment capability}.
\newblock \bibinfo{journal}{Journal of Statistical Physics}
  \bibinfo{volume}{175}, \bibinfo{pages}{47--70}.
\bibitem[{Kim et~al.(2016)Kim, Lee and Kim}]{Kim2016}
\bibinfo{author}{Kim, B.J.}, \bibinfo{author}{Lee, J.H.}, \bibinfo{author}{Kim,
  K.D.}, \bibinfo{year}{2016}.
\newblock \bibinfo{title}{Improvements of critical heat flux models for pool
  boiling on horizontal surfaces using interfacial instabilities of viscous
  potential flows}.
\newblock \bibinfo{journal}{International Journal of Heat and Mass Transfer}
  \bibinfo{volume}{93}, \bibinfo{pages}{200--206}.
\bibitem[{Kr\"{u}ger et~al.(2017)Kr\"{u}ger, Kusumaatmaja, Kuzmin, Shardt,
  Silva and Viggen}]{Kruger2017}
\bibinfo{author}{Kr\"{u}ger, T.}, \bibinfo{author}{Kusumaatmaja, H.},
  \bibinfo{author}{Kuzmin, A.}, \bibinfo{author}{Shardt, O.},
  \bibinfo{author}{Silva, G.}, \bibinfo{author}{Viggen, E.},
  \bibinfo{year}{2017}.
\newblock \bibinfo{title}{The Lattice Boltzmann Method}.
\newblock \bibinfo{publisher}{Springer International Publishing}.
\bibitem[{Kupershtokh et~al.(2009)Kupershtokh, Medvedev and
  Karpov}]{Kupershtokh_2009_equationsstatelattice}
\bibinfo{author}{Kupershtokh, A.L.}, \bibinfo{author}{Medvedev, D.A.},
  \bibinfo{author}{Karpov, D.I.}, \bibinfo{year}{2009}.
\newblock \bibinfo{title}{On equations of state in a lattice {B}oltzmann
  method}.
\newblock \bibinfo{journal}{Computers \& Mathematics with Applications}
  \bibinfo{volume}{58}, \bibinfo{pages}{965--974}.
\bibitem[{Lallemand and Luo(2000)}]{Lallemand_2000_TheorylatticeBoltzmann}
\bibinfo{author}{Lallemand, P.}, \bibinfo{author}{Luo, L.S.},
  \bibinfo{year}{2000}.
\newblock \bibinfo{title}{Theory of the lattice {B}oltzmann method: Dispersion,
  dissipation, isotropy, galilean invariance, and stability}.
\newblock \bibinfo{journal}{Physical Review E} \bibinfo{volume}{61},
  \bibinfo{pages}{6546}.
\bibitem[{LeVeque(2007)}]{Leveque2007}
\bibinfo{author}{LeVeque, R.}, \bibinfo{year}{2007}.
\newblock \bibinfo{title}{Finite Difference Methods for Ordinary and Partial
  Differential Equations}.
\newblock \bibinfo{publisher}{Society for Industrial and Applied Mathematics}.
\bibitem[{Li et~al.(2018a)Li, Huang and
  Kang}]{Li_2018_temperatureequationphase}
\bibinfo{author}{Li, Q.}, \bibinfo{author}{Huang, J.Y.}, \bibinfo{author}{Kang,
  Q.J.}, \bibinfo{year}{2018}a.
\newblock \bibinfo{title}{On the temperature equation in a phase change
  pseudopotential lattice {B}oltzmann model}.
\newblock \bibinfo{journal}{International Journal of Heat and Mass Transfer}
  \bibinfo{volume}{127}, \bibinfo{pages}{1112--1113}.
\bibitem[{Li et~al.(2015)Li, Kang, Francois, He and
  Luo}]{Li_2015_LatticeBoltzmannmodeling}
\bibinfo{author}{Li, Q.}, \bibinfo{author}{Kang, Q.J.},
  \bibinfo{author}{Francois, M.M.}, \bibinfo{author}{He, Y.L.},
  \bibinfo{author}{Luo, K.H.}, \bibinfo{year}{2015}.
\newblock \bibinfo{title}{Lattice {B}oltzmann modeling of boiling heat
  transfer: The boiling curve and the effects of wettability}.
\newblock \bibinfo{journal}{International Journal of Heat and Mass Transfer}
  \bibinfo{volume}{85}, \bibinfo{pages}{787--796}.
\bibitem[{Li and Luo(2013a)}]{Li2013}
\bibinfo{author}{Li, Q.}, \bibinfo{author}{Luo, K.}, \bibinfo{year}{2013}a.
\newblock \bibinfo{title}{Achieving tunable surface tension in the
  pseudopotential lattice boltzmann modeling of multiphase flows}.
\newblock \bibinfo{journal}{Physical Review E} \bibinfo{volume}{88}.
\bibitem[{Li et~al.(2014)Li, Luo, Kang and Chen}]{Li2014}
\bibinfo{author}{Li, Q.}, \bibinfo{author}{Luo, K.}, \bibinfo{author}{Kang,
  Q.}, \bibinfo{author}{Chen, Q.}, \bibinfo{year}{2014}.
\newblock \bibinfo{title}{Contact angles in the pseudopotential lattice
  boltzmann modeling of wetting}.
\newblock \bibinfo{journal}{Physical Review E} \bibinfo{volume}{90}.
\bibitem[{Li and Luo(2013b)}]{Li_2013_Achievingtunablesurface}
\bibinfo{author}{Li, Q.}, \bibinfo{author}{Luo, K.H.}, \bibinfo{year}{2013}b.
\newblock \bibinfo{title}{Achieving tunable surface tension in the
  pseudopotential lattice {B}oltzmann modeling of multiphase flows}.
\newblock \bibinfo{journal}{Physical Review E} \bibinfo{volume}{88},
  \bibinfo{pages}{053307}.
\bibitem[{Li et~al.(2016)Li, Luo, Kang, He, Chen and Liu}]{Li2016}
\bibinfo{author}{Li, Q.}, \bibinfo{author}{Luo, K.H.}, \bibinfo{author}{Kang,
  Q.J.}, \bibinfo{author}{He, Y.L.}, \bibinfo{author}{Chen, Q.},
  \bibinfo{author}{Liu, Q.}, \bibinfo{year}{2016}.
\newblock \bibinfo{title}{Lattice boltzmann methods for multiphase flow and
  phase-change heat transfer}.
\newblock \bibinfo{journal}{Progress in Energy and Combustion Science}
  \bibinfo{volume}{52}, \bibinfo{pages}{62--105}.
\bibitem[{Li et~al.(2013a)Li, Luo and Li}]{Li2013b}
\bibinfo{author}{Li, Q.}, \bibinfo{author}{Luo, K.H.}, \bibinfo{author}{Li,
  X.J.}, \bibinfo{year}{2013}a.
\newblock \bibinfo{title}{{Lattice Boltzmann modeling of multiphase flows at
  large density ratio with an improved pseudopotential model}}.
\newblock \bibinfo{journal}{Physical Review E} \bibinfo{volume}{87},
  \bibinfo{pages}{053301}.
\bibitem[{Li et~al.(2013b)Li, Luo and Li}]{Li_2013_LatticeBoltzmannmodeling}
\bibinfo{author}{Li, Q.}, \bibinfo{author}{Luo, K.H.}, \bibinfo{author}{Li,
  X.J.}, \bibinfo{year}{2013}b.
\newblock \bibinfo{title}{Lattice {B}oltzmann modeling of multiphase flows at
  large density ratio with an improved pseudopotential model}.
\newblock \bibinfo{journal}{Physical Review E} \bibinfo{volume}{87},
  \bibinfo{pages}{053301}.
\bibitem[{Li et~al.(2020a)Li, Yu and Wen}]{Li2020}
\bibinfo{author}{Li, Q.}, \bibinfo{author}{Yu, Y.}, \bibinfo{author}{Wen,
  Z.X.}, \bibinfo{year}{2020}a.
\newblock \bibinfo{title}{How does boiling occur in lattice boltzmann
  simulations?}
\newblock \bibinfo{journal}{Physics of Fluids} \bibinfo{volume}{32},
  \bibinfo{pages}{093306}.
\bibitem[{Li et~al.(2018b)Li, Yu, Zhou and Yan}]{Li2018}
\bibinfo{author}{Li, Q.}, \bibinfo{author}{Yu, Y.}, \bibinfo{author}{Zhou, P.},
  \bibinfo{author}{Yan, H.}, \bibinfo{year}{2018}b.
\newblock \bibinfo{title}{Enhancement of boiling heat transfer using
  hydrophilic-hydrophobic mixed surfaces: A lattice boltzmann study}.
\newblock \bibinfo{journal}{Applied Thermal Engineering} \bibinfo{volume}{132},
  \bibinfo{pages}{490--499}.
\bibitem[{Li et~al.(2021)Li, Li, Yu and Luo}]{Li2021}
\bibinfo{author}{Li, W.}, \bibinfo{author}{Li, Q.}, \bibinfo{author}{Yu, Y.},
  \bibinfo{author}{Luo, K.H.}, \bibinfo{year}{2021}.
\newblock \bibinfo{title}{{Nucleate boiling enhancement by structured surfaces
  with distributed wettability-modified regions: A lattice Boltzmann study}}.
\newblock \bibinfo{journal}{Applied Thermal Engineering} \bibinfo{volume}{194},
  \bibinfo{pages}{117130}.
\bibitem[{Li et~al.(2020b)Li, Li, Yu and Wen}]{Li2020a}
\bibinfo{author}{Li, W.}, \bibinfo{author}{Li, Q.}, \bibinfo{author}{Yu, Y.},
  \bibinfo{author}{Wen, Z.}, \bibinfo{year}{2020}b.
\newblock \bibinfo{title}{{Enhancement of nucleate boiling by combining the
  effects of surface structure and mixed wettability: A lattice Boltzmann
  study}}.
\newblock \bibinfo{journal}{Applied Thermal Engineering} \bibinfo{volume}{180},
  \bibinfo{pages}{115849}.
\bibitem[{Liang and Mudawar(2018)}]{Liang2018}
\bibinfo{author}{Liang, G.}, \bibinfo{author}{Mudawar, I.},
  \bibinfo{year}{2018}.
\newblock \bibinfo{title}{Pool boiling critical heat flux (chf) – part 1:
  Review of mechanisms, models, and correlations}.
\newblock \bibinfo{journal}{International Journal of Heat and Mass Transfer}
  \bibinfo{volume}{117}, \bibinfo{pages}{1352--1367}.
\bibitem[{Lienhard et~al.(1973)Lienhard, Dhir and Riherd}]{Lienhard1973}
\bibinfo{author}{Lienhard, J.H.}, \bibinfo{author}{Dhir, V.K.},
  \bibinfo{author}{Riherd, D.M.}, \bibinfo{year}{1973}.
\newblock \bibinfo{title}{{Peak Pool Boiling Heat-Flux Measurements on Finite
  Horizontal Flat Plates}}.
\newblock \bibinfo{journal}{Journal of Heat Transfer} \bibinfo{volume}{95},
  \bibinfo{pages}{477--482}.
\bibitem[{Liu et~al.(2016)Liu, Kang, Leonardi, Schmieschek, Narv{\'{a}}ez,
  Jones, Williams, Valocchi and Harting}]{Liu_2016_MultiphaselatticeBoltzmann}
\bibinfo{author}{Liu, H.}, \bibinfo{author}{Kang, Q.},
  \bibinfo{author}{Leonardi, C.R.}, \bibinfo{author}{Schmieschek, S.},
  \bibinfo{author}{Narv{\'{a}}ez, A.}, \bibinfo{author}{Jones, B.D.},
  \bibinfo{author}{Williams, J.R.}, \bibinfo{author}{Valocchi, A.J.},
  \bibinfo{author}{Harting, J.}, \bibinfo{year}{2016}.
\newblock \bibinfo{title}{{Multiphase lattice Boltzmann simulations for porous
  media applications}}.
\newblock \bibinfo{journal}{Computational Geosciences} \bibinfo{volume}{20},
  \bibinfo{pages}{777--805}.
\newblock \URLprefix \url{http://link.springer.com/10.1007/s10596-015-9542-3},
  \DOIprefix\doi{10.1007/s10596-015-9542-3}.
\bibitem[{Lou et~al.(2013)Lou, Guo and Shi}]{Lou2013}
\bibinfo{author}{Lou, Q.}, \bibinfo{author}{Guo, Z.}, \bibinfo{author}{Shi,
  B.}, \bibinfo{year}{2013}.
\newblock \bibinfo{title}{Evaluation of outflow boundary conditions for
  two-phase lattice boltzmann equation}.
\newblock \bibinfo{journal}{Physical Review E} \bibinfo{volume}{87}.
\bibitem[{Luo(1998)}]{Luo_1998_Unifiedtheorylattice}
\bibinfo{author}{Luo, L.S.}, \bibinfo{year}{1998}.
\newblock \bibinfo{title}{Unified theory of lattice {B}oltzmann models for
  nonideal gases}.
\newblock \bibinfo{journal}{Physical review letters} \bibinfo{volume}{81},
  \bibinfo{pages}{1618}.
\bibitem[{Lycett-Brown and Luo(2015)}]{LycettBrown_2015}
\bibinfo{author}{Lycett-Brown, D.}, \bibinfo{author}{Luo, K.H.},
  \bibinfo{year}{2015}.
\newblock \bibinfo{title}{{I}mproved forcing scheme in pseudopotential lattice
  {B}oltzmann methods for multiphase flow at arbitrarily high density ratios}.
\newblock \bibinfo{journal}{Physical Review E} \bibinfo{volume}{91},
  \bibinfo{pages}{23305}.
\newblock \DOIprefix\doi{10.1103/PhysRevE.91.023305}.
\bibitem[{Ma and Cheng(2019)}]{Ma_2019_3Dsimulationspool}
\bibinfo{author}{Ma, X.}, \bibinfo{author}{Cheng, P.}, \bibinfo{year}{2019}.
\newblock \bibinfo{title}{3d simulations of pool boiling above smooth
  horizontal heated surfaces by a phase-change lattice {B}oltzmann method}.
\newblock \bibinfo{journal}{International Journal of Heat and Mass Transfer}
  \bibinfo{volume}{131}, \bibinfo{pages}{1095--1108}.
\bibitem[{McCullough et~al.(2016)McCullough, Leonardi, Jones, Aminossadati and
  Williams}]{McCullough_2016_LatticeBoltzmannmethods}
\bibinfo{author}{McCullough, J.W.S.}, \bibinfo{author}{Leonardi, C.R.},
  \bibinfo{author}{Jones, B.D.}, \bibinfo{author}{Aminossadati, S.M.},
  \bibinfo{author}{Williams, J.R.}, \bibinfo{year}{2016}.
\newblock \bibinfo{title}{Lattice {B}oltzmann methods for the simulation of
  heat transfer in particle suspensions}.
\newblock \bibinfo{journal}{International Journal of Heat and Fluid Flow}
  \bibinfo{volume}{62}, \bibinfo{pages}{150--165}.
\bibitem[{Qian et~al.(1992)Qian, d'Humi{\`e}res and
  Lallemand}]{qian1992lattice}
\bibinfo{author}{Qian, Y.H.}, \bibinfo{author}{d'Humi{\`e}res, D.},
  \bibinfo{author}{Lallemand, P.}, \bibinfo{year}{1992}.
\newblock \bibinfo{title}{Lattice bgk models for navier-stokes equation}.
\newblock \bibinfo{journal}{EPL (Europhysics Letters)} \bibinfo{volume}{17},
  \bibinfo{pages}{479--484}.
\bibitem[{Safaei et~al.(2016)Safaei, Jahanbin, Kianifar, Gharehkhani, Kherbeet,
  Goodarzi and Dahari}]{Safaei_2016_MathematicalModelingNanofluids}
\bibinfo{author}{Safaei, M.R.}, \bibinfo{author}{Jahanbin, A.},
  \bibinfo{author}{Kianifar, A.}, \bibinfo{author}{Gharehkhani, S.},
  \bibinfo{author}{Kherbeet, A.S.}, \bibinfo{author}{Goodarzi, M.},
  \bibinfo{author}{Dahari, M.}, \bibinfo{year}{2016}.
\newblock \bibinfo{title}{Mathematical modeling for nanofluids simulation: A
  review of the latest works}, in: \bibinfo{booktitle}{Modeling and Simulation
  in Engineering Sciences}. \bibinfo{publisher}{{InTech}}.
\newblock \URLprefix \url{https://doi.org/10.5772/64154},
  \DOIprefix\doi{10.5772/64154}.
\bibitem[{Safari et~al.(2013)Safari, Rahimian and Krafczyk}]{Safari2013}
\bibinfo{author}{Safari, H.}, \bibinfo{author}{Rahimian, M.H.},
  \bibinfo{author}{Krafczyk, M.}, \bibinfo{year}{2013}.
\newblock \bibinfo{title}{{Extended lattice Boltzmann method for numerical
  simulation of thermal phase change in two-phase fluid flow}}.
\newblock \bibinfo{journal}{Physical Review E - Statistical, Nonlinear, and
  Soft Matter Physics} \bibinfo{volume}{88}, \bibinfo{pages}{1--12}.
\bibitem[{Sbragaglia et~al.(2007)Sbragaglia, Benzi, Biferale, Succi, Sugiyama
  and Toschi}]{Sbragaglia_2007_GeneralizedlatticeBoltzmann}
\bibinfo{author}{Sbragaglia, M.}, \bibinfo{author}{Benzi, R.},
  \bibinfo{author}{Biferale, L.}, \bibinfo{author}{Succi, S.},
  \bibinfo{author}{Sugiyama, K.}, \bibinfo{author}{Toschi, F.},
  \bibinfo{year}{2007}.
\newblock \bibinfo{title}{Generalized lattice {B}oltzmann method with
  multirange pseudopotential}.
\newblock \bibinfo{journal}{Physical Review E} \bibinfo{volume}{75},
  \bibinfo{pages}{026702}.
\bibitem[{Shan and Chen(1993)}]{ShanChen1993}
\bibinfo{author}{Shan, X.}, \bibinfo{author}{Chen, H.}, \bibinfo{year}{1993}.
\newblock \bibinfo{title}{Lattice {B}oltzmann model for simulating flows with
  multiple phases and components}.
\newblock \bibinfo{journal}{Physical Review E} \bibinfo{volume}{47},
  \bibinfo{pages}{1815--1819}.
\bibitem[{Shan and He(1998)}]{Shan_1998_Discretizationvelocityspace}
\bibinfo{author}{Shan, X.}, \bibinfo{author}{He, X.}, \bibinfo{year}{1998}.
\newblock \bibinfo{title}{Discretization of the velocity space in the solution
  of the {B}oltzmann equation}.
\newblock \bibinfo{journal}{Physical Review Letters} \bibinfo{volume}{80},
  \bibinfo{pages}{65}.
\bibitem[{Son et~al.(1999)Son, Dhir and Ramanujapu}]{Son1999}
\bibinfo{author}{Son, G.}, \bibinfo{author}{Dhir, V.K.},
  \bibinfo{author}{Ramanujapu, N.}, \bibinfo{year}{1999}.
\newblock \bibinfo{title}{{Dynamics and Heat Transfer Associated With a Single
  Bubble During Nucleate Boiling on a Horizontal Surface}}.
\newblock \bibinfo{journal}{Journal of Heat Transfer} \bibinfo{volume}{121},
  \bibinfo{pages}{623--631}.
\bibitem[{Son et~al.(2002)Son, Ramanujapu and Dhir}]{son2002numerical}
\bibinfo{author}{Son, G.}, \bibinfo{author}{Ramanujapu, N.},
  \bibinfo{author}{Dhir, V.K.}, \bibinfo{year}{2002}.
\newblock \bibinfo{title}{Numerical simulation of bubble merger process on a
  single nucleation site during pool nucleate boiling}.
\newblock \bibinfo{journal}{J. Heat Transfer} \bibinfo{volume}{124},
  \bibinfo{pages}{51--62}.
\bibitem[{Sun et~al.(2022)Sun, Li, Li, Wen and Liu}]{Sun2022}
\bibinfo{author}{Sun, X.}, \bibinfo{author}{Li, Q.}, \bibinfo{author}{Li, W.},
  \bibinfo{author}{Wen, Z.}, \bibinfo{author}{Liu, B.}, \bibinfo{year}{2022}.
\newblock \bibinfo{title}{{Enhanced pool boiling on microstructured surfaces
  with spatially-controlled mixed wettability}}.
\newblock \bibinfo{journal}{International Journal of Heat and Mass Transfer}
  \bibinfo{volume}{183}, \bibinfo{pages}{122164}.
\bibitem[{Swift et~al.(1995)Swift, Osborn and
  Yeomans}]{Swift_1995_LatticeBoltzmannSimulation}
\bibinfo{author}{Swift, M.R.}, \bibinfo{author}{Osborn, W.R.},
  \bibinfo{author}{Yeomans, J.M.}, \bibinfo{year}{1995}.
\newblock \bibinfo{title}{Lattice {B}oltzmann simulation of nonideal fluids}.
\newblock \bibinfo{journal}{Physical Review Letters} \bibinfo{volume}{75},
  \bibinfo{pages}{830--833}.
\bibitem[{Wang and Dhir(1993)}]{10.1115/1.2910737}
\bibinfo{author}{Wang, C.H.}, \bibinfo{author}{Dhir, V.K.},
  \bibinfo{year}{1993}.
\newblock \bibinfo{title}{{Effect of Surface Wettability on Active Nucleation
  Site Density During Pool Boiling of Water on a Vertical Surface}}.
\newblock \bibinfo{journal}{Journal of Heat Transfer} \bibinfo{volume}{115},
  \bibinfo{pages}{659--669}.
\bibitem[{Welch and Wilson(2000)}]{Welch2000}
\bibinfo{author}{Welch, W.J.}, \bibinfo{author}{Wilson, J.},
  \bibinfo{year}{2000}.
\newblock \bibinfo{title}{{A Volume of Fluid Based Method for Fluid Flows with
  Phase Change}}.
\newblock \bibinfo{journal}{Journal of Computational Physics}
  \bibinfo{volume}{160}, \bibinfo{pages}{662--682}.
\bibitem[{Yu et~al.(2005)Yu, Girimaji and Luo}]{Yu_2005_DNSLESdecaying}
\bibinfo{author}{Yu, H.}, \bibinfo{author}{Girimaji, S.S.},
  \bibinfo{author}{Luo, L.S.}, \bibinfo{year}{2005}.
\newblock \bibinfo{title}{{DNS} and {LES} of decaying isotropic turbulence with
  and without frame rotation using lattice {B}oltzmann method}.
\newblock \bibinfo{journal}{Journal of Computational Physics}
  \bibinfo{volume}{209}, \bibinfo{pages}{599--616}.
\bibitem[{Yuan and Schaefer(2006)}]{Yuan_2006_Equationsstatelattice}
\bibinfo{author}{Yuan, P.}, \bibinfo{author}{Schaefer, L.},
  \bibinfo{year}{2006}.
\newblock \bibinfo{title}{Equations of state in a lattice {B}oltzmannn model}.
\newblock \bibinfo{journal}{Physics of Fluids} \bibinfo{volume}{18},
  \bibinfo{pages}{042101}.
\bibitem[{Zhai et~al.(2017)Zhai, Zheng and
  Zheng}]{Zhai_2017_PseudopotentiallatticeBoltzmann}
\bibinfo{author}{Zhai, Q.}, \bibinfo{author}{Zheng, L.},
  \bibinfo{author}{Zheng, S.}, \bibinfo{year}{2017}.
\newblock \bibinfo{title}{Pseudopotential lattice {B}oltzmann equation method
  for two-phase flow: A higher-order chapmann-enskog expansion}.
\newblock \bibinfo{journal}{Physical Review E} \bibinfo{volume}{95},
  \bibinfo{pages}{023313}.
\bibitem[{Zhang and Chen(2003)}]{Zhang2003}
\bibinfo{author}{Zhang, R.}, \bibinfo{author}{Chen, H.}, \bibinfo{year}{2003}.
\newblock \bibinfo{title}{Lattice boltzmann method for simulations of
  liquid-vapor thermal flows}.
\newblock \bibinfo{journal}{Phys. Rev. E} \bibinfo{volume}{67},
  \bibinfo{pages}{066711}.
\bibitem[{Zhao et~al.(2021)Zhao, Liang, Sun and Wang}]{Zhao2021}
\bibinfo{author}{Zhao, W.}, \bibinfo{author}{Liang, J.}, \bibinfo{author}{Sun,
  M.}, \bibinfo{author}{Wang, Z.}, \bibinfo{year}{2021}.
\newblock \bibinfo{title}{{Investigation on the effect of convective outflow
  boundary condition on the bubbles growth, rising and breakup dynamics of
  nucleate boiling}}.
\newblock \bibinfo{journal}{International Journal of Thermal Sciences}
  \bibinfo{volume}{167}, \bibinfo{pages}{106877}.
\bibitem[{Zheng et~al.(2019)Zheng, Zheng and Zhai}]{Zheng2019}
\bibinfo{author}{Zheng, L.}, \bibinfo{author}{Zheng, S.},
  \bibinfo{author}{Zhai, Q.}, \bibinfo{year}{2019}.
\newblock \bibinfo{title}{{Analysis of force treatment in lattice Boltzmann
  equation method}}.
\newblock \bibinfo{journal}{International Journal of Heat and Mass Transfer}
  \bibinfo{volume}{139}, \bibinfo{pages}{747--750}.
\newblock \URLprefix
  \url{https://doi.org/10.1016/j.ijheatmasstransfer.2019.05.059},
  \DOIprefix\doi{10.1016/j.ijheatmasstransfer.2019.05.059}.
\bibitem[{Zou and He(1997)}]{Zou1997}
\bibinfo{author}{Zou, Q.}, \bibinfo{author}{He, X.}, \bibinfo{year}{1997}.
\newblock \bibinfo{title}{On pressure and velocity boundary conditions for the
  lattice boltzmann {BGK} model}.
\newblock \bibinfo{journal}{Physics of Fluids} \bibinfo{volume}{9},
  \bibinfo{pages}{1591--1598}.

\end{thebibliography}
\end{document}